\documentclass[aos, preprint]{imsart}

\RequirePackage[OT1]{fontenc}
\RequirePackage{amsthm,amsmath}
\usepackage{array}
\usepackage{mathrsfs}
\usepackage{amssymb,bbm}
\usepackage{amsfonts}
\usepackage{amsmath,amssymb}
\usepackage{graphicx}
\usepackage{amscd}
\usepackage{color}
\usepackage{mathrsfs}
\usepackage{latexsym,bm}
\usepackage{array}
\usepackage{threeparttable}

\usepackage[plain,noend]{algorithm2e}

\DeclareFontFamily{U}{mathx}{\hyphenchar\font45}
\DeclareFontShape{U}{mathx}{m}{n}{
      <5> <6> <7> <8> <9> <10>
      <10.95> <12> <14.4> <17.28> <20.74> <24.88>
      mathx10
      }{}
\DeclareSymbolFont{mathx}{U}{mathx}{m}{n}
\DeclareMathAccent{\widecheck}{\mathalpha}{mathx}{"71}

\DeclareMathOperator*{\plim}{plim}

\startlocaldefs \numberwithin{equation}{section} \theoremstyle{it}
\newtheorem{thm}{Theorem}[section]

\newtheorem{ass}{Assumption}[section]
\newtheorem{cor}{Corollary}[section]
\newtheorem{pro}{Proposition}[section]
\newtheorem{rem}{Remark}
\endlocaldefs

\newcommand{\e}{\varepsilon}

\begin{document}

\begin{frontmatter}
\title{Statistical inference for autoregressive models under heteroscedasticity of unknown form}
\runtitle{ Inference for AR models}

\begin{aug}
\author{\fnms{Ke} \snm{Zhu}\thanksref{t1}
\ead[label=e1]{mazhuke@hku.hk}}

\thankstext{t1}{Supported in part by National Natural Science Foundation of China (No.11571348, 11371354, 11690014, 11731015 and 71532013). This paper
was initiated when the author worked at the Chinese Academy of Sciences.}
\runauthor{K. Zhu}

\affiliation{University of Hong Kong}
\address{University of Hong Kong\\
Department of Statistics and Actuarial Science\\
Pok Fu Lam Road\\
Hong Kong\\
\printead{e1}}
\end{aug}

\begin{abstract}
This paper provides an entire inference procedure for the autoregressive model
under (conditional) heteroscedasticity of unknown form with a finite variance.
We first establish the asymptotic normality of the weighted least absolute deviations estimator
(LADE) for the model. Second, we develop the random weighting (RW) method to estimate its
asymptotic covariance matrix,
leading to the implementation of the Wald test. Third, we construct a portmanteau test for model checking, and use
the RW method to obtain its critical values. As a special weighted LADE, the feasible adaptive LADE (ALADE) is proposed and
proved to have the same efficiency as its infeasible counterpart.
The importance of our entire methodology based on the feasible ALADE is illustrated by simulation results and the real data
analysis on three U.S. economic data sets.
\end{abstract}



\end{frontmatter}

\section{Introduction}
Consider a $p$-th order heteroscedastic  auto-regressive (AR) model:
\begin{flalign}
y_{t}&=\mu_0+\phi_{10}y_{t-1}+\cdots+\phi_{p0}y_{t-p}+\e_{t},\label{1.1}
\end{flalign}
where the error $\e_{t}$ satisfies
\begin{flalign}
\e_{t}&=g_{t}u_{t}:=g\left(\frac{t}{n}\right)u_{t}\label{1.2}
\end{flalign}
for $t=1,\cdots,n$. Here,
$g(\cdot)$ is a positive bounded scalar function with unknown form, $u_{t}$ is a re-scaled error with unknown form  and a finite variance for all $t$, and $n$
is the sample size.
In model (\ref{1.1}), $y_{t}$ is allowed to be non-stationary with a finite variance
due to the heteroscedasticity of $\e_{t}$, but it can not be a unit root process, since
the coefficients \{$\phi_{i0}\}_{i=1}^{p}$ are required to
satisfy the stationarity condition of AR($p$) model. In model (\ref{1.2}), $\{u_{t}\}$
need not be independent and identically distributed (i.i.d.), nor
even a martingale difference sequence (m.d.s.), allowing for model mis-specification.
This setting is important, in view of that the literature generally needs a non-i.i.d. sequence $\{u_{t}\}$ (see, e.g., Drost and Nijman (1993)).
Under certain identification condition on $u_t$, $g_t$ is
identifiable. Particularly, when $u_{t}$ is stationary, the (conditionally) heteroscedastic structure of $\e_{t}$ is allowed to change
abruptly, gradually, or periodically according to the un-specified form of $g(\cdot)$. Similar specifications as for $g_{t}$ have been widely adopted in the literature; see, e.g.,
Robinson (1989, 2012), Fryzlewicz, Sapatinas, and Subba Rao (2006), Zhou and Wu (2009),  and Chen and Hong (2016)
to name a few.



Model (\ref{1.1}) is one of the often used models of empirical macroeconomics and statistics.
Conventional statistical inference methods for this model are designed for homogenous error $\{\e_{t}\}$ (i.e., $E\e_{t}^{2}=\mbox{a constant}$),
which is either a sequence of i.i.d. random variables or an m.d.s.;
see, e.g.,
Gon\c{c}alves and Kilian (2004), Yao and Brockwell (2006), Zhu and Ling (2015) and references therein.
However, homogenous error
can be a crucial weakness in applications, since heteroscedasticity has been widely demonstrated
by Tsay (1988) for social science data, Watson (1999) for interest rates data,
Busetti and Taylor (2003) and Sensier and van Dijk (2004) for macroeconomic data,
 Amado and Ter\"{a}svirta (2013, 2014) for stock index data, and many others.
As shown in Diebold (1986), Mikosch and
St\v{a}ric\v{a} (2004), and St\v{a}ric\v{a} and Granger (2005), the presence of heteroscedasticity could mislead the conventional
time series analysis procedure resulting in erroneous conclusions. Hence, it is necessary to develop a
valid statistical inference procedure for model (\ref{1.1}) under heteroscedasticity.

So far few works have centered around this topic, and most of them are based on the least squares estimator (LSE); see
Nicholls and Pagan (1983)
for earlier works and Phillips and Xu (2006) for more recent ones.
The seminal work in Carroll (1982) and Robinson (1987) demonstrated that the LSE in regression models is less efficient than the adaptive LSE (ALSE), which takes the unknown heteroscedastic form of the error term into account; see also Cragg (1983) for the study of
the instrumental variable LSE.
Motivated by this,
Xu and Phillips (2008) constructed a feasible ALSE for model (\ref{1.1}), and showed that this feasible ALSE is more efficient than the LSE. However, their theory does not allow $u_{t}$ to be conditionally heteroscedastic, and their methodology may be limited in practice since
no valid statistical inference tools (e.g., $t$-test, Wald test and model diagnostic checking) is provided.
Moreover, a heavy-tailed error is an often observed feature in fitted model (\ref{1.1}), and
the LSE-based inference methods in this case are undesirable from the viewpoint of robustness.


This paper provides an entire statistical inference procedure for model (\ref{1.1}) based on
the  weighted least absolute deviations estimator (LADE).
We first establish the asymptotic normality of
the weighted LADE. Second, since the asymptotic covariance matrix of the weighted LADE can not be estimated directly from the sample, we develop the random weighting (RW) method to estimate this covariance matrix,  leading to the implementation of
the Wald test. The RW method
initiated by Jin, Ying, and Wei (2001)
has been widely applied to provide statistical
inference; see Zhu (2016). Third, we construct a portmanteau test for model diagnostic checking, and use the RW method to obtain its critical values.
As a special weighted LADE, the infeasible adaptive LADE (ALADE) with weight equal to $g_{t}$ is desired in terms of efficiency, and its
advantage in efficiency is demonstrated by a formal comparison.
To circumvent the un-observable $g_{t}$, we further propose a feasible ALADE, prove that
it has the same efficiency as its infeasible counterpart, and
show that our entire inference procedure is still valid based on this feasible estimator.
Under the identification condition $E|u_{t}|=1$ for $g_{t}$,
this feasible ALADE is constructed with weight equal to a kernel estimator of $g_{t}$.
Unlike
Xu and Phillips (2008),
our entire inference procedure based on the weighted LADE
does not rule out the conditional heteroscedasticity of $u_{t}$.
The importance of our methodology is illustrated by simulation results and the real data
analysis on three U.S. economic data sets.

We emphasize that the idea of the weighted LADE has been adopted for the linear regression model
with heteroscedasticity of either known form in Gutenbrunner and Jure\v{c}kov\'{a} (1992) or unknown form
in Zhao (2001). But our technique is different from theirs, since they neither considered a general heteroscedastic time series model as models (\ref{1.1})-(\ref{1.2}), nor
provided an entire valid inference procedure.
When $\e_{t}$ is stationary and conditionally heteroscedastic with unknown form,
Zhu and Ling (2015) provided the statistical inference method for model (\ref{1.1}) based on
either the self-weighted LADE or the usual LADE. Our weighted LADE and their self-weighted LADE are not compatible due to different
weighting mechanisms. The weight $w_{sw,t}$ for the self-weighted LADE is a random variable in the filtration generated by
 $\{y_{k}\}_{k\leq t-1}$, and it is useful to deal with the infinite variance AR model.
 For the finite variance AR model (as in our setting), $w_{sw,t}$ is not needed any more,
 since the self-weighted
 LADE is always less efficient than the usual LADE in this case.
On the contrary,
 the weight $w_{t}$
 for our weighted LADE is a deterministic function (or its kernel estimate based on the whole sample $\{y_{t}\}_{t=1}^{n}$), and it is
 used to take the unknown heteroscedastic form of $\e_{t}$ into account. As a result, the weighted LADE
can be more efficient than the usual LADE by choosing an appropriate weight
(e.g., the weight for the feasible ALADE) in many circumstances. Compared to the usual LADE in Zhu and Ling (2015),
our weighted LADE has two more incremental contributions besides the advantage in efficiency.
First, the results based on the weighted LADE allow $\e_{t}$ to be heteroscedastic with unknown form, and hence the scope of applications is much wider in dealing with the finite variance AR model.
Second, although the RW method is motivated by Zhu and Ling (2015),
new proof techniques are used to handle the heteroscedasticity of $\e_{t}$,
especially for the feasible ALADE.

The remainder of the paper is organized as follows. Section 2
obtains the asymptotic normality of the weighted LADE.
Section 3 presents the RW method to estimate the asymptotic covariance matrix.
Section 4 constructs a portmanteau test for model diagnostic checking.
Section 5 considers the choice of the weight function, and discusses the asymptotic efficiency.
Section 6 proposes the feasible ALADE. Simulation results are reported in Section 7.
Concluding remarks are offered in Section 8.
Proofs of all theorems are relegated to Appendixes. Additional simulation results, applications, some technical lemmas and the remaining proofs
are provided in the supplementary material (Zhu, 2018).

Throughout the paper, $|\cdot|$ and $\|\cdot\|$ denote the absolute value and the vector $l_{2}$-norm, respectively,
$A'$ is the transpose of matrix $A$, $\|\xi\|_{\kappa}=(E|\xi|^{\kappa})^{1/\kappa}$ is the $L^{\kappa}$-norm of random variable $\xi$,
$\plim$ denotes the convergence in probability, $\to_{d}$ denotes the convergence in distribution, $I(\cdot)$ is the indicator function,
and $\mbox{sgn}(a)=I(a>0)-I(a<0)$ is the sign of any $a\in\mathcal{R}$.

\section{The weighted LADE}

Let $\theta=(\mu,\phi_1,\cdots,\phi_p)'\in\Theta$ be the unknown parameter of model (\ref{1.1}),
and $\theta_0=(\mu_0,\phi_{10},\cdots,\phi_{p0})'\in\Theta$ be its true value, where $\Theta$ is the parameter space,
and it is a compact subset of  $\mathcal{R}^{p+1}$ with $\mathcal{R}=(-\infty,\infty)$.
Given the observations $\{y_{t}\}_{t=1}^{n}$ and the
initial values $\{y_{t}\}_{t=-p}^{0}\equiv 0$, we consider the weighted least absolute deviations estimator (LADE) for $\theta_{0}$:
\begin{flalign*}
\widehat{\theta}_{wn}=\arg\min_{\theta\in\Theta}\sum_{t=1}^{n}\frac{\left|y_{t}-Y_{t-1}'\theta\right|}{w_{t}},
\end{flalign*}
where $Y_{t-1}=(1,y_{t-1},\cdots,y_{t-p})'$ and $w_{t}:=w(t/n)$ are weights with $w(\cdot)$ being a positive scalar function.
Particularly, when $w(\cdot)\equiv 1$, $\widehat{\theta}_{wn}$ becomes the usual LADE, denoted by $\widehat{\theta}_{n}$.
To study the asymptotic theory of $\widehat{\theta}_{wn}$, we need the notation of
near-epoch dependent (NED) random variables.

\vspace{3mm}

{\sc Definition 1.} {\it\,\,
$\{Z_{nt}\}$ is near-epoch dependent in $L^{\kappa}$-norm ($L^{\kappa}$-NED) on
$\{V_{nt}\}$ if
$\|Z_{nt}-E(Z_{nt}|\mathfrak{F}_{t-m}^{t+m})\|_{\kappa}\leq d_{nt}\psi_{m}$, where
$\mathfrak{F}_{t-m}^{t+m}:=\sigma(V_{ns}; t-m\leq s\leq t+m)$ is a filtration with $\sigma(\cdot)$ being the $\sigma$-field operator,
$\{d_{nt}\}$ are positive constants, and $\psi_{m}\to0$ as $m\to\infty$.
}

\vspace{3mm}

\noindent The definition of near-epoch dependence was first introduced in Billingsley (1968) and has been widely used
in the literature.
NED processes allow for considerable heterogeneity and also for dependence and include the mixing processes as a special case.
Many non-linear models are shown to be NED (see, e.g., Davidson (2002, 2004) for overviews), and
hence the NED concept makes possible a convenient theory of inference for these models.
Besides near-epoch dependence, the physical dependence in Wu (2005) is a parallel tool to depict the dependence
of a dynamic process. It might be interesting to apply  physical dependence to our setting, and we leave this for future study.

Let
$f_{t}(x)$ be the conditional density of $u_{t}$ given $\mathcal{F}_{t-1}$, where
$\mathcal{F}_{t}:=\sigma(u_{s}; s\leq t)$ is a filtration. We make the following five assumptions throughout the paper, where
the definition of $f_{t}(x)$ is essential only for the last one.

\begin{ass}
$\theta_0$ is an interior point in $\Theta$ and for each
$\theta\in\Theta$, $\phi(z)\not=0$ when $|z|\leq 1$, where $\phi(z):=1-\sum_{i=1}^p \phi_i z^i$. \label{a2.1}
\end{ass}

\begin{ass}
$g(\cdot)$ and $w(\cdot)$ are continuous
on the interval $[0, 1]$ except at a finite number of points,  and
\begin{flalign*}
& 0<\underline{C}\leq\inf_{x\in[0,1]}g(x)
\leq\sup_{x\in[0,1]}g(x)\leq\overline{C}<\infty,\\
& 0<\underline{C}\leq\inf_{x\in[0,1]}w(x)
\leq\sup_{x\in[0,1]}w(x)\leq\overline{C}<\infty,
\end{flalign*}
\noindent for some positive
numbers $\underline{C}$ and $\overline{C}$.\label{a2.2}
\end{ass}

\begin{ass}
$E[\mbox{sgn}(\e_{t})|\mathcal{F}_{t-1}]=0$.\label{a2.3}
\end{ass}

\begin{ass}
$u_{t}$ is an $\alpha$-mixing but not necessary a stationary process, and

(i) $\sup_{t}E|u_{t}|^{4+\delta_{0}}<\infty$ for some $\delta_{0}>0$;

(ii) $m_{r}^{(1)}:=E(u_{t-r})$ is uniformly bounded for all $t, r\geq1$;

(iii) $m_{r,s}^{(2)}:=E(u_{t-r}u_{t-s})$ is uniformly bounded for all $t, r, s\geq1$.
\label{a2.4}
\end{ass}

\begin{ass}
$f_{t}(x)$ satisfies that

(i) $f_{t}(x)$ is continuous with $f_{t}(0)>0$ and $\sup_{x\in\mathcal{R}}f_{t}(x)<\infty$ (a.s.) for all $t\geq1$;

(ii) $\{f_{t}(0)\}$ is $L^{2+\delta_{1}}$-NED on $\{u_{t}\}$ for some $\delta_{1}>0$;

(iii) $\tau_{0}:=E[f_{t}(0)]$ is uniformly bounded for all $t\geq1$;

(iv) $\tau_{r}^{(1)}:=E[f_{t}(0)u_{t-r}]$ is uniformly bounded for all $t, r\geq1$;

(v) $\tau_{r,s}^{(2)}:=E[f_{t}(0)u_{t-r}u_{t-s}]$ is uniformly bounded for all $t, r, s\geq1$.\label{a2.5}
\end{ass}

We offer some remarks on the aforementioned assumptions. Assumption \ref{a2.1} is the usual stationarity condition for AR($p$) model, and it implies that
\begin{flalign}
y_{t}=\rho+\sum_{i=0}^{\infty}\alpha_{i}\e_{t-i},\label{2.1}
\end{flalign}
where $\rho=\mu_0/(1-\phi_{10}-\cdots-\phi_{p0})$ and
$\sum_{i=0}^{\infty}|\alpha_i|<\infty$.
Assumption \ref{a2.2} is sufficient to guarantee that both
$g(\cdot)$ and $w(\cdot)$ are integrable on the interval $[0, 1]$ up to any finite order, and similar conditions have been
used in Cavaliere (2004), Cavaliere and Taylor (2007), Phillips and Xu (2006), Xu and Phillips (2008), and many others.
Assumption \ref{a2.3}
is the identification condition for $\theta_{0}$ based on the weighted LADE.
Assumption \ref{a2.4} is adopted from Kuersteiner (2002) and Gon\c{c}alves and Kilian (2004)  in a similar way.
Assumption \ref{a2.4}(i) is stronger than the condition that $Eu_{t}^{2}<\infty$, which is
sufficient for the asymptotic normality of the LADE if $\e_{t}$ is stationary; see, e.g.,
Zhu and Ling (2015) and references therein. We resort to a stronger moment condition
of $u_{t}$ in Assumption \ref{a2.4}(i) to handle the heteroscedasticity of $\e_{t}$.
Assumption \ref{a2.4}(ii)-(iii) are less restrictive than the condition that
$u_{t}$ is covariance stationary with mean zero. Assumption
\ref{a2.5} is required only for the LADE but not for the LSE, and it is stronger than the one in Bassett and Koenker (1978),
Davis and Dunsmuir (1997), Pan, Wang, and Yao (2007), and Zhu and Ling (2012, 2015)
due to the heteroscedasticity of $\e_{t}$. When $f_{t}(0)\equiv f(0)$ is independent of $t$, Assumption \ref{a2.5}(ii)-(iii) hold automatically, and
Assumption \ref{a2.5}(iv)-(v) become Assumption \ref{a2.4}(ii)-(iii), respectively.
When $u_{t}$ follows a stationary autoregressive conditional heteroskedasticity (ARCH) type
 model, the NED condition in Assumption
\ref{a2.5}(ii) can be checked or even be un-needed, and the moment conditions in
Assumption
\ref{a2.5}(iii)-(v) can be satisfied (see the discussions for Corollaries 2.1-2.3 below).

We shall point out that Assumption \ref{a2.4}(ii)-(iii) and Assumption \ref{a2.5}(iii)-(v) need
$m_{r}^{(1)}$, $m_{r,s}^{(2)}$, $\tau_{0}$, $\tau_{r}^{(1)}$ and $\tau_{r,s}^{(2)}$ to be independent of $t$.
In the general framework, these terms may depend on $t$.
A detailed study for this general framework might be done in a separate paper.


Let $d_{1}=\int_{0}^{1}\frac{1}{w^{2}(x)}dx$,
$d_{2}=\tau_{0}\int_{0}^{1}\frac{1}{g(x)w(x)}dx$, and
\begin{flalign}
\zeta_{r}^{(1)}&=\rho d_{1}+\left(\sum_{i=0}^{\infty}\alpha_{i}m_{i+r}^{(1)}\right)\int_{0}^{1}\frac{g(x)}{w^{2}(x)}dx,\label{2.2}\\
\zeta_{r}^{(2)}&=\rho d_{2}+\left(\sum_{i=0}^{\infty}\alpha_{i}\tau_{i+r}^{(1)}\right)\int_{0}^{1}\frac{1}{w(x)}dx,\label{2.3}\\
\xi_{r,s}^{(1)}&=\rho\left[\zeta_{r}^{(1)}+\zeta_{s}^{(1)}\right]-\rho^{2}d_{1}
+\left(\sum_{i=0}^{\infty}\sum_{j=0}^{\infty}\alpha_{i}\alpha_{j}
m_{i+r,j+s}^{(2)}\right)\int_{0}^{1}\frac{g^{2}(x)}{w^{2}(x)}dx,\label{2.4}\\
\xi_{r,s}^{(2)}&=\rho\left[\zeta_{r}^{(2)}+\zeta_{s}^{(2)}\right]-\rho^{2}d_{2}
+\left(\sum_{i=0}^{\infty}\sum_{j=0}^{\infty}\alpha_{i}\alpha_{j}
\tau_{i+r,j+s}^{(2)}\right)\int_{0}^{1}\frac{g(x)}{w(x)}dx.\label{2.5}
\end{flalign}
With these notations, define
\begin{flalign}
\Sigma_{l}=
\left(
\begin{array}{cc}
d_{l} & K_{l}'\\
K_{l} & \Omega_{l}
\end{array}
\right)\in\mathcal{R}^{(p+1)\times(p+1)}
\,\,\mbox{ for }\,\,
l=1, 2,\label{2.6}
\end{flalign}
where
$K_{l}$ is the $p\times 1$ column vector with $r$-th element $\zeta_{r}^{(l)}$, and
$\Omega_{l}$ is the $p\times p$ matrix with $(r,s)$-th element $\xi_{r,s}^{(l)}$. Our first main result is
given as follows:

\begin{thm}\label{thm1}
Suppose Assumptions \ref{a2.1}-\ref{a2.5} hold. Then,
$$\sqrt{n}(\widehat{\theta}_{wn}-\theta_{0})\to_{d}N\left(0,
\frac{1}{4}\Sigma_{2}^{-1}\Sigma_{1}\Sigma_{2}^{-1}\right)$$
as $n\to\infty$, where $\Sigma_{l}$ ($l=1, 2$) is defined as in (\ref{2.6}).
\end{thm}

\begin{rem}
Theorem \ref{thm1} implies that $\widehat{\theta}_{wn}$ is
asymptotically normal for general specifications of $g(\cdot)$ and $u_{t}$,
which determine the asymptotic variance of $\widehat{\theta}_{wn}$, and hence need
be considered in all inferential methods. Since there is no guarantee that the structures of $g(\cdot)$ and
$u_{t}$ specified by the researcher provide a correct description of reality, we will propose valid
inferential methods without specifying the forms of $g(\cdot)$ and $u_{t}$ in the sequel.

When $g(\cdot)\equiv 1$, $w(\cdot)\equiv 1$, and $u_{t}$ is  stationary, Theorem \ref{thm1} covers
the existing results on the usual LADE in Bassett and Koenker (1978) for the i.i.d. $u_{t}$
and Zhu and Ling (2015) for the conditionally heteroscedastic $u_{t}$ with unknown form.


When $\mu_0\equiv0$ (i.e., $\rho\equiv0$) and $f_{t}(0)\equiv f(0)$, the asymptotic variance of $\widehat{\theta}_{wn}$ has the following simplified version:
\begin{flalign}\label{simple_version_1}
\frac{1}{4}\Sigma_{2}^{-1}\Sigma_{1}\Sigma_{2}^{-1}=\frac{c_{gw}}{4f^2(0)}\Lambda^{-1}
,
\end{flalign}
where $c_{gw}=\big(\int_{0}^{1} \frac{g^{2}(x)}{w^{2}(x)} dx\big)
\big(\int_{0}^{1} \frac{g(x)}{w(x)} dx\big)^{-2}$
and $\Lambda$ is the $p\times  p$ matrix with $(r, s)$-th element
$\sum_{i=0}^{\infty}\sum_{j=0}^{\infty}\alpha_{i}\alpha_{j}
m_{i+r,j+s}^{(2)}$. Furthermore, if $u_{t}$ is a sequence of i.i.d. random variables,
the  $(r, s)$-th element of $\Lambda$ becomes $\lambda_{|r-s|}$, where
$\lambda_{k}=Eu_{t}^2(\sum_{i=0}^{\infty}\alpha_{i}\alpha_{i+k})$.
\end{rem}

\begin{rem}
Our weighted LADE $\widehat{\theta}_{wn}$ requires $E[\mbox{sgn}(\e_{t})|\mathcal{F}_{t-1}]=0$, based on which
model (\ref{1.1}) gives the predicted median of future $y_t$ under $L_1$ loss function. If one is interested to predict the mean of future
$y_t$, the prediction  can be conducted under $L_2$ loss function, assuming $E[\e_{t}|\mathcal{F}_{t-1}]=0$.
From the viewpoint of robustness, the predicted median of future $y_t$ under $L_1$ loss function may be desired, and if $E[\mbox{sgn}(\e_{t})|\mathcal{F}_{t-1}]=E[\e_{t}|\mathcal{F}_{t-1}]=0$, it also
forecasts the mean of future $y_t$.

We also note that though $\widehat{\theta}_{wn}$ has the parametric convergence rate $n^{-1/2}$
in Theorem \ref{thm1}, it is not needed for the purpose of prediction.
 A less restricted model (\ref{1.1}) with time-varying AR coefficients might be studied in the future to delivery a better prediction of $y_t$.
\end{rem}

In the finite variance AR model, the technique in Zhu and Ling (2015) for the usual LADE requires the stationarity of $y_t$,
which is against our setting
for a either time-varying $g_t$ or non-stationary $u_t$. Our proof techniques for the weighted LADE in Theorem \ref{thm1} or
its related inferential methods below
rely on Theorem 1 of Andrews (1988), and they require a key technical condition that
$\{f_{t}(0)\}$ is $L^{2+\delta_1}$-NED on $\{u_t\}$ (see Assumption \ref{a2.5}(ii)), which seems new to the study of the LADE.
Moreover, in the present of $g_t$, we further propose a special weighted LADE with $w_t$ equal to a kernel estimator of $g_t$; see Section 6 below.
In many situations, this special weighted LADE is ``optimal'' in terms of efficiency, but its related proof techniques (particularly for
 Propositions A.3-A.4 in Appendix A) are new and not available in
Zhu and Ling (2015).

In the infinite variance AR model, Zhu and Ling (2015) constructed a self-weighted LADE,
which is $\sqrt{n}$-consistent and asymptotically normal when $y_t$ is stationary.
Meanwhile, they pointed out that the convergence rate of
the usual LADE is slower than $n^{-1/2}$ when $\e_{t}$ follows the first order
(G)ARCH model, while Davis (1996) proved that the convergence rate of
the usual LADE is faster than $n^{-1/2}$ when $\e_{t}$ is i.i.d.
This implies that the self-weighted LADE may not always outperform the usual LADE (or the weighted LADE) when $y_t$ is stationary with an infinite variance. In general, when $y_t$ is allowed to be non-stationary (as in our setting) with an infinite variance,
the asymptotic properties of the weighted LADE and the self-weighted LADE are not clear at this stage, and
we leave this topic for future study.

Next, we relax the technical conditions in Theorem \ref{thm1} for some special cases of model (\ref{1.1}).
We first consider the case that $u_{t}$ has the  ARCH-type structure (Engle, 1982; Bollerslev, 1986; Giraitis, Kokoska, and Leipus, 2000; Francq and Zako\"{i}an, 2010):
\begin{flalign}
u_{t}=\sigma_{t}\eta_{t},\label{2.7}
\end{flalign}
where $u_{t}$ is stationary, $\{\eta_{t}\}$ is a sequence of i.i.d. innovations with median zero, and $\sigma_{t}\in\mathcal{F}_{t-1}$ satisfies that
$\sigma_{t}\geq \underline{c}$ for some $\underline{c}>0$. This condition on $\sigma_{t}$ directly holds for most of
ARCH-type models as long as their intercept term has a positive lower bound.
Since $u_{t}$ is stationary,  Assumption \ref{a2.4}(ii)-(iii) are not needed any more if Assumption \ref{a2.4}(i) holds.


Let $f_{\eta}(\cdot)$ be the density of $\eta_{t}$. Under model (\ref{2.7}), $f_{t}(0)=f_{\eta}(0)/\sigma_{t}$.
Since $\sigma_t\geq\underline{c}$, Assumption \ref{a2.5}(ii) holds  if
$\{\sigma_t\}$ is $L^{2+\delta_{1}}$-NED on $\{u_{t}\}$, and Assumption \ref{a2.5}(iii)-(v)
hold by Assumption \ref{a2.4}(i).
Hence, Assumption \ref{a2.5} can be relaxed to Assumption \ref{a2.6} in this case.

\begin{ass}
$u_{t}$ satisfies model (\ref{2.7}), and

(i) $f_{\eta}(x)$ is continuous with $f_{\eta}(0)>0$ and $\sup_{x\in\mathcal{R}}f_{\eta}(x)<\infty$;

(ii) $\{\sigma_{t}\}$ is $L^{2+\delta_{1}}$-NED on $\{u_{t}\}$ for some $\delta_{1}>0$.\label{a2.6}
\end{ass}

\begin{cor}\label{cor2.1}
Suppose Assumptions \ref{a2.1}-\ref{a2.3}, \ref{a2.4}(i) and \ref{a2.6} hold. Then,
the result in Theorem \ref{thm1} holds.
\end{cor}

\noindent Since $\sigma_t\geq\underline{c}$, by a minor extension of lemma 2.1 of Davidson (2002), we can show that
Assumption \ref{a2.6}(ii) holds if
$\{\sigma_{t}^{\kappa_1}\}$ is $L^{2+\delta_{1}}$-NED on $\{u_{t}\}$ for some $\kappa_1\geq 1$.
When $\sigma_{t}$ in (\ref{2.7}) satisfies that
\begin{flalign}\label{ARCH_infinity}
\sigma_{t}^{\kappa_1}=\psi_0+\sum_{i=1}^{\infty}\psi_{i} |u_{t-i}|^{\kappa_1}
\end{flalign}
with $\psi_0>0$ and $\psi_{i}\geq 0$ for all $i\geq 1$,
it is straightforward to see that $\{\sigma_{t}^{\kappa_1}\}$ is $L^{\kappa_2}$-NED (for some $\kappa_2\geq1$) on $\{u_{t}\}$
if
\begin{flalign}\label{NED_condition}
\sum_{i=1}^{\infty}\psi_i<\infty\,\,\,\mbox{ and }\,\,\,E|u_{t}|^{\kappa_1\kappa_2}<\infty.
\end{flalign}
Hence, Assumption \ref{a2.6}(ii) holds
under (\ref{ARCH_infinity})-(\ref{NED_condition}) with $\kappa_2=2+\delta_1$, which allow a general class of
ARCH-type models, including (G)ARCH, ARCH($\infty$) (Robinson, 1991), power GARCH (Ding, Granger, and Engle, 1993), hyperbolic GARCH (Davidson, 2004),  and many others. Note that our NED condition
in (\ref{NED_condition}) is different from the one in Davidson (2004), since our filtration is based on
 $\{u_{t}\}$ instead of $\{\eta_{t}\}$.


Second, we consider the case that $f_{t}(0)$ depends on finite lags of $u_{t}$. In this case,
Assumption \ref{a2.4}(i) and Assumption \ref{a2.5}(ii) can be further relieved as shown in Assumption \ref{a2.7}, which
does not need $\{f_{t}(0)\}$ to be NED.

\begin{ass}
(i) $\sup_{t}E|u_{t}|^{2+\delta_{0}}<\infty$ for some $\delta_{0}>0$; (ii) $f_{t}(0)\in\sigma(u_{s};t-c_{0}\leq s\leq t)$
for some $c_{0}>0$ is uniformly bounded for all $t\geq1$.\label{a2.7}
\end{ass}

\begin{cor}\label{cor2.2}
Suppose Assumptions \ref{a2.1}-\ref{a2.3}, \ref{a2.4}(ii)-(iii), \ref{a2.5}(i) and (iii)-(v), and \ref{a2.7} hold. Then,
the result in Theorem \ref{thm1} holds.
\end{cor}

\noindent Clearly, Assumption \ref{a2.7}(ii) holds under model (\ref{ARCH_infinity}) with
$\psi_i\equiv 0$ for all $i\geq i_0$ and some integer  $i_0\geq 1$.
This implies that when $u_{t}$ follows a stationary finite-lag ARCH model with $E|u_{t}|^{2+\delta_{0}}<\infty$ and $Eu_{t}^{4}=\infty$,
$\widehat{\theta}_{wn}$ is asymptotically normal. However,
the LSE in this case is asymptotically non-normal with a slower convergence rate than $n^{-1/2}$
as shown in Zhang and Ling (2015). Hence, the weighted LADE seems to be more convenient and efficient than the LSE
to tackle the heavy-tailedness of $\e_{t}$.

Third, we consider the case that $\phi_{i0}\equiv 0$, under which model (\ref{1.1}) becomes a
location model. For this location model, Assumptions \ref{a2.1} is redundant, and
Assumptions \ref{a2.4}-\ref{a2.5} can be replaced by Assumption \ref{a2.8} below.

\begin{ass}
$\phi_{i0}\equiv 0$ for all $i$ in model (\ref{1.1}), and either

(i) Assumption \ref{a2.5}(i) and (iii) hold and $\{f_{t}(0)\}$ is $L^{1}$-NED on $\{u_{t}\}$; or

(ii) Assumption \ref{a2.6}(i) holds and
$\{\sigma_{t}\}$ is $L^{1}$-NED on $\{u_{t}\}$ under (\ref{2.7}).\label{a2.8}
\end{ass}

\begin{cor}\label{cor2.3}
Suppose Assumptions \ref{a2.2}-\ref{a2.3}
and \ref{a2.8} hold. Then,
the result in Theorem \ref{thm1} holds.
\end{cor}

\noindent As argued before, Assumption \ref{a2.8}(ii) holds under (\ref{ARCH_infinity})-(\ref{NED_condition}) with $\kappa_2=1$.
It is worth noting that the weaker technical assumptions in Corollaries \ref{cor2.1}-\ref{cor2.3}
not only suffice for Theorem \ref{thm1} but also for other asymptotic results subsequently.

\section{The bootstrapped variance estimator}

Since the asymptotic variance of $\widehat{\theta}_{wn}$ in Theorem \ref{thm1} depends on the
unknown forms of $g(\cdot)$ and $u_{t}$, it can not be estimated directly by
its sample counterpart.  To solve this problem, we use the random weighting (RW)
method to approximate the limiting distribution of $\widehat{\theta}_{wn}$ in Theorem \ref{thm1}.
Let $\{w^{*}_{t}\}_{t=1}^{n}$
be a sequence of i.i.d. nonnegative random variables, with mean and variance both equal to 1. Define
\begin{flalign}
\widehat{\theta}_{wn}^{*}=\arg\min_{\theta\in\Theta}\sum_{t=1}^{n}w_{t}^{*}\frac{\left|y_{t}-Y_{t-1}'\theta\right|}{w_{t}}.
\end{flalign}
Based on Assumption \ref{a3.1}, we can show that the distribution of
$\sqrt{n}(\widehat{\theta}_{wn}-\theta_{0})$ can be approximated
by the resampling distribution of $\sqrt{n}(\widehat{\theta}_{wn}^{*}-\widehat{\theta}_{wn})$.

\begin{ass}\label{a3.1}
(i) $E(w_{t}^{*})^{2+\delta_{2}}<\infty$
for some $\delta_{2}>0$; (ii)
$\{w_{t}^{*}\}$ and $\{y_{t}\}$
are independent.
\end{ass}

\begin{thm}\label{thm2}
Suppose Assumption \ref{a3.1} and the conditions in Theorem \ref{thm1} hold. Then, conditional
on $\{y_{t}\}_{t=1}^{n}$,
\begin{flalign*}
\sqrt{n}(\widehat{\theta}_{wn}^{*}-\widehat{\theta}_{wn})\to_{d}N\left(0,
\frac{1}{4}\Sigma_{2}^{-1}\Sigma_{1}\Sigma_{2}^{-1}\right)\,\,\,\,\mbox{ in probability}
\end{flalign*}
as $n\to\infty$, where $\Sigma_{l}$ ($l=1, 2$) is defined as in (\ref{2.6}).
\end{thm}

\begin{rem}
The RW method can be viewed as a variant of the wild bootstrap
in Wu (1986) and Liu (1988).
It provides us a valid tool to implement the statistical inference based on the weighted LADE, and
some other bootstrap methods (see, e.g., Gon\c{c}alves and Kilian, 2004)
may also be valid or even better under our setting. An interesting work is to find the optimal bootstrap method among all valid ones in theory.
This will necessitate higher order asymptotic analysis and is beyond the scope of this paper.
\end{rem}

According to Theorems \ref{thm1} and \ref{thm2}, we can approximate the
asymptotic covariance matrix of $\widehat{\theta}_{wn}$ by the following procedure:

1.  Generate $J$ replications of the i.i.d. random weights $\{w_{t}^{*}\}_{t=1}^{n}$
from the standard exponential distribution, which has mean and variance both equal
to one.

2. Compute $\widehat{\theta}_{wn}^{*}$ for $i$-th replication, and denote it as $b_{i}$.

3. Calculate the sample variance-covariance matrix of $\{b_{i}-\widehat{\theta}_{wn}\}_{i=1}^{J}$, denoted by
$\widehat{V}_{wn}$, which provides a good approximation for the asymptotic covariance matrix of
$\widehat{\theta}_{wn}$ for large $J$.

Based on $\widehat{V}_{wn}$, we can construct a
Wald test statistic
\begin{flalign}\label{3.2}
W_{wn}=(\Gamma\widehat{\theta}_{wn}-r)'(\Gamma\widehat{V}_{wn}\Gamma')^{-1}(\Gamma\widehat{\theta}_{wn}-r)
\end{flalign}
to test  the following linear null hypothesis:
\begin{flalign}\label{3.3}
H_{0}: \Gamma\theta_{0}=r,
\end{flalign}
where $\Gamma$ is an $s\times (p+1)$
constant matrix with rank
$s$ and $r$ is an $s\times1$
constant vector. If $W_{wn}$
is larger than the upper-tailed critical value of $\chi^{2}_{s}$, the null hypothesis $H_{0}$
is rejected. Otherwise, $H_{0}$
is not rejected.

\section{Model diagnostic checking}

Model diagnostic checking is an important step in applications.
Most of the existing methods, including the time domain correlation-based tests
and frequency domain periodogram-based tests, require that the observed data are stationary with
a homogenous innovation; see, e.g., Hong (1996), Hong and Lee (2005), Escanciano (2006), and
Zhu and Li (2015) for an overview.
Hence, they are invalid for model (\ref{1.1}), which calls for a new method for
its diagnostic checking. In this section, we construct a sign-based Ljung-Box
portmanteau test as in Zhu and Ling (2015) for this purpose.

Let $\e_{t}(\theta)=y_{t}-\theta'Y_{t-1}$. The idea of our sign-based portmanteau test is based on the fact that
$\{\mbox{sgn}(\e_{t}(\theta_{0}))\}$ is a sequence of un-correlated random variables under Assumption \ref{a2.3}. Hence,
if model (\ref{1.1}) is correctly specified, it is expected that
the sample
autocorrelation function of
$\{\mbox{sgn}(\e_{t}(\widehat{\theta}_{wn}))\}$ at lag $k$, denoted by $\widehat{r}_{wn, k}$, is close to zero, where
$$\widehat{r}_{wn,k}=\frac{\sum_{t=k+1}^{n}\left[
\mbox{sgn}(\e_{t}(\widehat{\theta}_{wn}))-\overline{\e}(\widehat{\theta}_{wn})\right]
\left[
\mbox{sgn}(\e_{t-k}(\widehat{\theta}_{wn}))-\overline{\e}(\widehat{\theta}_{wn})\right]}
{\sum_{t=1}^{n}\left[
\mbox{sgn}(\e_{t}(\widehat{\theta}_{wn}))-\overline{\e}(\widehat{\theta}_{wn})\right]^2}$$
with $\overline{\e}(\theta)=\frac{1}{n}\sum_{t=1}^{n}\mbox{sgn}(\e_{t}(\theta))$.
Let $\widehat{r}_{wn}=(\widehat{r}_{wn,1},\widehat{r}_{wn,2},\cdots,\widehat{r}_{wn,M})'$ for some integer $M\geq1$.
To study the joint distribution of $\widehat{r}_{wn}$, we need one more assumption below.

\begin{ass}\label{a4.1}
$u_{t}$ and $f_{t}(0)$ satisfy that

(i) $m_{r,s}^{(3)}:=E\left[\mbox{sgn}(u_{t-r})u_{t-s}\right]$ is uniformly bounded for all $t, s, r\geq1$;


(ii) $\mu_{r}^{(1)}:=E[f_{t}(0)\mbox{sgn}(u_{t-r})]$ is uniformly bounded for all $t, r\geq1$;

(iii) $\mu_{r,s}^{(2)}:=E[f_{t}(0)\mbox{sgn}(u_{t-r})u_{t-s}]$ is uniformly bounded for all $t, s, r\geq1$.
\end{ass}


Let $d_{3}=\int_{0}^{1}\frac{1}{g(x)}dx$ and
$$
K_{3k}=
\left(\begin{array}{c}
0\\
\sum_{i=0}^{\infty}\alpha_{i}m_{k,1+i}^{(3)}\\
\vdots\\
\sum_{i=0}^{\infty}\alpha_{i}m_{k,p+i}^{(3)}
\end{array}\right),
\,\,\,\,
\Omega_{3k}=
\left(\begin{array}{c}
d_{3}\mu_{k}^{(1)}\\
\rho d_{3}\mu_{k}^{(1)}+\sum_{i=0}^{\infty}\alpha_{i}\mu_{k,1+i}^{(2)}\\
\vdots\\
\rho d_{3}\mu_{k}^{(1)}+\sum_{i=0}^{\infty}\alpha_{i}\mu_{k,p+i}^{(2)}
\end{array}\right)
$$
for $k=1,\cdots,M$. With these notations, define
\begin{flalign}\label{4.1}
\Sigma_{3}=
\left(\begin{array}{cc}
I_{M} & K_{3}' \\
K_{3} & \Sigma_{1}
\end{array}\right)
\in\mathcal{R}^{\overline{p}\times \overline{p}}\,\,\mbox{ and }\,\,
\Sigma_{4}=(I_{M}, -\Omega_{3}\Sigma_{2}^{-1})\in\mathcal{R}^{M\times \overline{p}},
\end{flalign}
where $K_{3}=\big(\int_{0}^{1}\frac{g(x)}{w(x)}dx\big)\dot{K}_{3}$ with $\dot{K}_{3}=(K_{31},K_{32},\cdots,K_{3M})\in\mathcal{R}^{(p+1)\times M}$,
$\Omega_{3}=(\Omega_{31},\Omega_{32},\cdots,\Omega_{3M})'
\in\mathcal{R}^{M\times (p+1)}$, and $\overline{p}=p+M+1$.
The following theorem gives the joint distribution of $\widehat{r}_{wn}$.

\begin{thm}\label{thm3}
Suppose Assumption \ref{a4.1} and the conditions of Theorem \ref{thm1} hold. Then, if model
(\ref{1.1}) is correctly specified,
$$\sqrt{n}\widehat{r}_{wn}\to_{d} N(0,\Sigma_{4}\Sigma_{3}\Sigma_{4}')$$
as $n\to\infty$, where $\Sigma_{l}\,(l=3, 4)$ is defined as in (\ref{4.1}).
\end{thm}

\begin{rem}
When $g(\cdot)\equiv 1$, $w(\cdot)\equiv 1$, and $u_{t}$ is  stationary, Theorem \ref{thm3} covers
the existing result on the usual LADE in Zhu and Ling (2015) for the conditionally heteroscedastic $u_{t}$ with unknown form.

When $\mu_0\equiv0$ and $f_{t}(0)\equiv f(0)$, the asymptotic variance of $\widehat{r}_{wn}$ has the following simplified version:
\begin{flalign}\label{simple_version_2}
\Sigma_{4}\Sigma_{3}\Sigma_{4}'=I_{M}-(2-c_{gw})\ddot{K}_{3}'\Lambda^{-1}\ddot{K}_{3},
\end{flalign}
where $\ddot{K}_{3}$ is the sub-matrix of
$\dot{K}_{3}$ by removing its first row, and $c_{gw}$ and $\Lambda$ are defined as in (\ref{simple_version_1}).
\end{rem}

Since the forms of $g(\cdot)$ and $u_{t}$ are not specified,
the asymptotic covariance matrix of $\widehat{r}_{wn}$ can not be directly estimated from its
sample counterpart.
We use the similar RW method as in Section 3 to avoid this difficulty.
Define
$$\widehat{r}_{wn,k}^{*}=\frac{\sum_{t=k+1}^{n}w_{t}^{*}\left[
\mbox{sgn}(\e_{t}(\widehat{\theta}_{wn}^{*}))-\overline{\e}(\widehat{\theta}_{wn}^{*})\right]
\left[
\mbox{sgn}(\e_{t-k}(\widehat{\theta}_{wn}^{*}))-\overline{\e}(\widehat{\theta}_{wn}^{*})\right]}
{\sum_{t=1}^{n}\left[
\mbox{sgn}(\e_{t}(\widehat{\theta}_{wn}^{*}))-\overline{\e}(\widehat{\theta}_{wn}^{*})\right]^2}.$$
Let $\widehat{r}_{wn}^{*}=(\widehat{r}_{wn,1}^{*},\widehat{r}_{wn,2}^{*},\cdots,\widehat{r}_{wn,M}^{*})'$.
The following theorem indicates that  the distribution of
$\sqrt{n}\widehat{r}_{wn}$ can be approximated
by the resampling distribution of $\sqrt{n}(\widehat{r}_{wn}^{*}-\widehat{r}_{wn})$.

\begin{thm}\label{thm4}
Suppose Assumption \ref{a3.1} and the conditions in Theorem \ref{thm3} hold. Then, if model
(\ref{1.1}) is correctly specified, conditional
on $\{y_{t}\}_{t=1}^{n}$,
\begin{flalign*}
\sqrt{n}(\widehat{r}_{wn}^{*}-\widehat{r}_{wn})\to_{d}N\left(0,
\Sigma_{4}\Sigma_{3}\Sigma_{4}'\right)\,\,\,\,\mbox{ in probability}
\end{flalign*}
as $n\to\infty$, where $\Sigma_{l}$ ($l=3, 4$) is defined as in (\ref{4.1}).
\end{thm}

According to Theorems \ref{thm3} and \ref{thm4}, we can approximate the
asymptotic covariance matrix of $\widehat{r}_{wn}$ by the following procedure:

1.  Generate $J$ replications of the i.i.d. random weights $\{w_{t}^{*}\}_{t=1}^{n}$
from the standard exponential distribution.

2. Compute $\widehat{r}_{wn}^{*}$ for $i$-th replication, and denote it as $c_{i}$.

3. Calculate the sample variance-covariance matrix of $\{c_{i}-\widehat{r}_{wn}\}_{i=1}^{J}$, denoted by
$\widehat{U}_{wn}$, which provides a good approximation for the asymptotic covariance matrix of
$\widehat{r}_{wn}$ for large $J$.

Based on $\widehat{U}_{wn}$, we can construct a
portmanteau test statistic
\begin{flalign}\label{4.3}
S_{wn}(M)=\widehat{r}_{wn}'\widehat{U}_{wn}^{-1}\widehat{r}_{wn}
\end{flalign}
to detect  the adequacy of model (\ref{1.1}). If $S_{wn}(M)$
is larger than the upper-tailed critical value of $\chi^{2}_{M}$, the fitted model (\ref{1.1})
is not adequate. Otherwise, it is adequate.

\section{The choice of weight function}

By choosing the weight function $w_{t}\equiv1$, a systematical statistical inference procedure for model (\ref{1.1}) based on the usual LADE $\widehat{\theta}_{n}$ is already
available in Sections 2-4.  However, this choice of the weight function may lead to a deficient weighted LADE
 in terms of efficiency. In this section, we are interested in finding the ``optimal'' weight $w_{t}$ to minimize the asymptotic covariance matrix in Theorem 2.1.
Given $w_{t}=g_{t}$, we consider a weighted LADE  defined by
\begin{flalign*}
\widehat{\theta}_{an}=\arg\min_{\theta\in\Theta}\sum_{t=1}^{n}\frac{\left|y_{t}-Y_{t-1}'\theta\right|}{g_{t}}.
\end{flalign*}
Clearly, $\widehat{\theta}_{an}$ is infeasible in practice as $g_{t}$ is not observable.
However, Corollary \ref{cor5.1} below shows that $\widehat{\theta}_{an}$ is the desired one
 under three different scenarios.

\begin{cor}\label{cor5.1}
$\frac{1}{4}\Sigma_{2}^{-1}\Sigma_{1}\Sigma_{2}^{-1}$ attains its minimum $A$ when $w_{t}=g_{t}$ (up to a
constant multiplier) under one of the following scenarios:

(S1) $\mu_0\equiv 0$;

(S2) $f_{t}(0)\equiv f(0)$;

(S3) $\phi_{i0}\equiv0$ for all $i$.

\noindent Particularly, under (S1),  $A$ is the $p\times p$ matrix with $(r,s)$-th element $a_{r,s}$, and
$a_{r,s}=\frac{1}{4}[a_{r,s}^{(\tau)}]^{-1}a_{r,s}^{(m)}[a_{r,s}^{(\tau)}]^{-1}$ with
$a_{r,s}^{(\tau)}=\sum_{i=0}^{\infty}\sum_{j=0}^{\infty}\alpha_{i}\alpha_{j}
\tau_{i+r,j+s}^{(2)}$ and $a_{r,s}^{(m)}=\sum_{i=0}^{\infty}\sum_{j=0}^{\infty}\alpha_{i}\alpha_{j}
m_{i+r,j+s}^{(2)}$.
\end{cor}

Under (S1), the asymptotic covariance matrix of
$\widehat{\theta}_{an}$ is adaptive to the unknown form of $g(\cdot)$.
From this point of view, we follow Xu and Philips (2008) to call $\widehat{\theta}_{an}$ the infeasible adaptive LADE (ALADE).
Obviously, Corollary \ref{cor5.1} indicates that $\widehat{\theta}_{an}$ is more efficient than $\widehat{\theta}_{n}$
under (S1), (S2) or (S3). Our finding that $g_{t}$ is the optimal weight is
similar to those in
Gutenbrunner and Jure\v{c}kov\'{a} (1992) and Zhao (2001) for the linear heteroscedastic regression model.


Next, we make a formal comparison of efficiency among the LADE $\widehat{\theta}_{n}$,
the infeasible ALADE $\widehat{\theta}_{an}$, the LSE $\widecheck{\theta}_{n}$ in Phillips and Xu (2006),
and the infeasible adaptive LSE (ALSE) $\widecheck{\theta}_{an}$ in Xu and Phillips (2008),
where
\begin{flalign*}
\widecheck{\theta}_{n}&=\left(\sum_{t=1}^{n}Y_{t-1}Y_{t-1}'\right)^{-1}\left(\sum_{t=1}^{n}y_{t}Y_{t-1}\right)\\
&\,\,\mbox{ and }\,\,
\widecheck{\theta}_{an}=\left(\sum_{t=1}^{n}g_{t}^{-2}Y_{t-1}Y_{t-1}'\right)^{-1}\left(\sum_{t=1}^{n}g_{t}^{-2}y_{t}Y_{t-1}\right).
\end{flalign*}
We assume that $\mu_0\equiv0$ as in Xu and Phillips (2008),   and
that $\{u_{t}\}$ is an i.i.d. sequence with $\mbox{median}(u_{t})=0$, $Eu_{t}=0$, and $Eu_{t}^{2}=1$
to make the comparison feasible.  As $\{u_{t}\}$ is an i.i.d. sequence,
the moment condition that $E|u_{t}|^{2+\delta_{0}}<\infty$ for some $\delta_{0}>0$ is sufficient for the asymptotic normality of
$\widehat{\theta}_{n}$ and $\widehat{\theta}_{an}$ by Corollary 2.2, and this is also the case for
$\widecheck{\theta}_{n}$ and $\widecheck{\theta}_{an}$ by a slight modification of the proof in Xu and Philips (2008).

Define
\begin{flalign*}
b_{1}=\frac{1}{4f^{2}(0)}\frac{\int_{0}^{1}g^{2}(x)dx}{\left[\int_{0}^{1}g(x)dx\right]^{2}},\,\,
b_{2}=\frac{1}{4f^{2}(0)},\,\, b_{3}=\frac{\int_{0}^{1}g^{4}(x)dx}{\left[\int_{0}^{1}g^{2}(x)dx\right]^{2}},\,\,
\mbox{ and }\,\,b_{4}=1,
\end{flalign*}
where $f(\cdot)$ is the density function of $u_{t}$. Under the aforementioned conditions,
result (\ref{simple_version_1}) implies that as $n\to\infty$,
\begin{flalign*}
\sqrt{n}(\widehat{\theta}_{n}-\theta_{0})\to_{d}
N\left(0,b_{1}\Lambda^{-1}\right)\,\,\mbox{ and }\,\,
\sqrt{n}(\widehat{\theta}_{an}-\theta_{0})\to_{d}
N\left(0,b_{2}\Lambda^{-1}\right),
\end{flalign*}
respectively; and Phillips and Xu (2006) and Xu and Phillips (2008) showed that as $n\to\infty$,
\begin{flalign*}
\sqrt{n}(\widecheck{\theta}_{n}-\theta_{0})\to_{d}
N\left(0,b_{3}\Lambda^{-1}\right)\,\,\mbox{ and }\,\,
\sqrt{n}(\widecheck{\theta}_{an}-\theta_{0})\to_{d}
N\left(0,b_{4}\Lambda^{-1}\right),
\end{flalign*}
respectively. The asymptotic efficiencies of these four estimators can be formally compared by just looking at the values of
$b_{i}$ under different scenarios of
$u_{t}$ and $g(\cdot)$.
Below, Example 1 considers the case that $g(\cdot)$ has an abrupt change in the variance, when
$u_{t}$ is chosen to be an i.i.d. $\mbox{standardized }t_{3}$ ($\mbox{ST}_{3}$) or $\mbox{standardized Laplace}(0,1)$ ($\mbox{SL}(0,1)$)
sequence with $E(u_{t}^{2})=1$. For the cases that $g(\cdot)$ has gradual and periodical change in the variance,
one can refer to Examples 2 and 3 in Zhu (2018).

\vspace{3mm}

{\sc Example 1.}
({\it An abrupt change in the variance})
Let $\tau\in[0,1]$ and $g(\cdot)$ be the step function
\begin{flalign}
g(x)=e_{0}+(e_{1}-e_{0})I(x\geq \tau),\label{2.9}
\end{flalign}
where $s\in[0,1]$, $e_{0}>0$, and $e_{1}>0$.
Under (\ref{2.9}), the variance of $\e_{t}$ is $e_{0}^{2}$ before the break point $[n\tau]$, and
$e_{1}^{2}$ afterwards. Let $\delta=e_{1}/e_{0}$. Then,  some algebra shows that
\begin{flalign*}
b_{1}=\frac{1}{4f^{2}(0)}\frac{\tau+(1-\tau)\delta^{2}}{(\tau+(1-\tau)\delta)^2}\,\,\mbox{ and }\,\,
b_{3}=\frac{\tau+(1-\tau)\delta^{4}}{(\tau+(1-\tau)\delta^{2})^2}.
\end{flalign*}
Fig\,\ref{fig1}(a)-(f) plot the values of all $b_{i}$ in terms of $\delta$
for $\tau=0.1, 0.5$, and 0.9, respectively.  From this figure, our findings are as follows:

\begin{figure}[!h]
\centering$
\begin{array}{c}
\includegraphics[width=5in,height=4in]{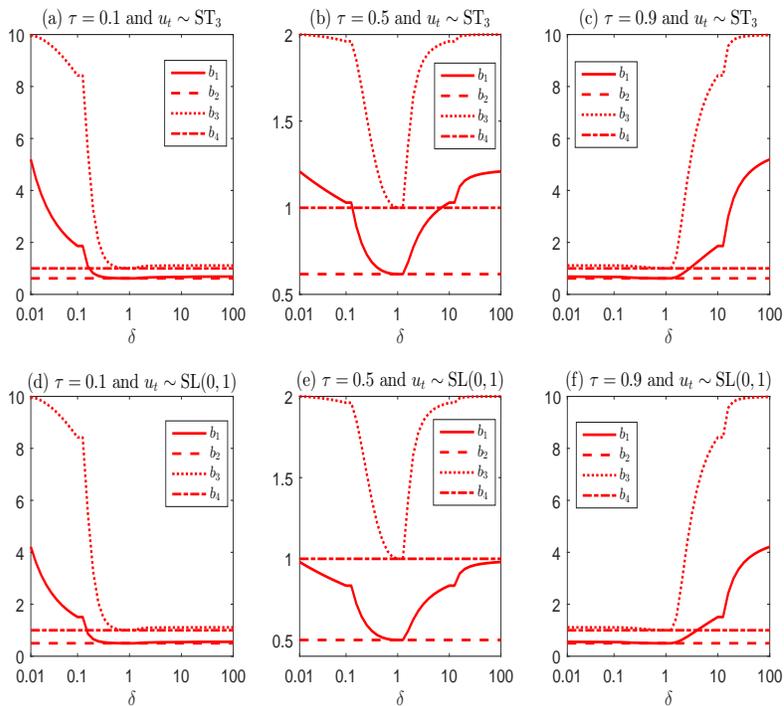}
\end{array}$
\caption{The values of $b_{1}$ (solid line),  $b_{2}$ (dashed line), $b_{3}$ (dash-dot line), and $b_{4}$ (dotted line) across $\delta$}\label{fig1}
\end{figure}

(i) $\widehat{\theta}_{an}$ is more efficient than $\widecheck{\theta}_{an}$ under all considered cases. Meanwhile, $\widehat{\theta}_{n}$ can be more efficient than $\widecheck{\theta}_{an}$ when the break happens earlier with a large $\delta$,
the break happens later with a small $\delta$, or the break happens in the middle with $\delta$ not largely deviating from 1.
All of these findings imply the efficiency advantage of the ALADE (or LADE) over the ALSE when
$u_{t}$ is heavy-tailed.

(ii) $\widehat{\theta}_{an}$ (or $\widecheck{\theta}_{an}$) is
always more efficient than $\widehat{\theta}_{n}$ (or $\widecheck{\theta}_{n}$) as expected, but the efficiency advantage is less significant when
the break happens earlier with a large $\delta$, the break happens later with a small $\delta$, or the break happens in the middle
with $\delta\approx 1$ (see, Xu and Philips (2008, p.270) for a similar phenomenon and explanation).
This indicates that the ALADE with an appropriate choice of the weight function can have
a pronouncing efficiency advantage over the usual LADE.

Overall, Examples 1-3 demonstrate that
(i) $\widehat{\theta}_{an}$ and $\widehat{\theta}_{n}$
can be more efficient than $\widecheck{\theta}_{an}$ when $u_{t}$ is heavy-tailed; (ii)
$\widehat{\theta}_{an}$ is more efficient than $\widehat{\theta}_{n}$ in all considered cases.
However, $\widehat{\theta}_{an}$ is infeasible in practice, and this problem will be solved in the next section.

\section{The feasible adaptive LADE}
The objective of this section is to give a feasible ALADE,
prove that this feasible ALADE calculated with estimated weights is equivalent to
the infeasible ALADE $\widehat{\theta}_{an}$ asymptotically, and
show that our entire statistical inference procedure in Sections 2-4 is still valid based on this feasible ALADE.

We use a similar  non-parametric method as in
Xu and Phillips (2008) to achieve this goal. Let $\widehat{\e}_{t}=y_{t}-Y_{t-1}'\widehat{\theta}_{n}$ be the residual from the LADE. Denote
\begin{flalign}
\widehat{g}_{t}=\sum_{i=1}^{n}k_{ti}|\widehat{\e}_{i}|,\label{6.1}
\end{flalign}
where $k_{ti}=(\sum_{i=1}^{n}K_{ti})^{-1}K_{ti}$ with
\begin{flalign*}
K_{ti}=
\left\{\begin{array}{ll}
K\left(\frac{t-i}{nb}\right) & \mbox{ if }t\not=i,\\
0 & \mbox{ if }t=i,
\end{array}\right.
\end{flalign*}
where $K(\cdot)$ is the kernel function defined on $\mathcal{R}$, and $b>0$ is the bandwidth.
Of course, the role of
$\widehat{g}_{t}$ is to deputize for $g_{t}$, which is the mean of $|\e_{t}|$ based on the following assumption:

\begin{ass}
$E|u_{t}|=1$.\label{a6.1}
\end{ass}

\noindent
Assumption \ref{a6.1} is used for the identification of $g_{t}$, and it does not
rule out the conditional heteroscedasticity of $u_{t}$. For the feasible ALSE,
Xu and Phillips (2008) resorted to the identification condition $E[u_{t}^{2}|\mathcal{F}_{t-1}]=1$;
however, their identification condition is restrictive, since it does not allow $u_{t}$ to be conditionally heteroscedastic.
Moreover, we shall mention that
besides (\ref{6.1}), other methods could also be used to estimate $g_{t}$ (see, e.g., Fan and Yao, 1998; Yu and Jones, 2004), and
this is a promising direction for the future study.

By using $\widehat{g}_{t}$ in (\ref{6.1}), our feasible ALADE is defined as follows:
\begin{flalign*}
\widetilde{\theta}_{an}:=\arg\min_{\theta\in\Theta}\sum_{t=1}^{n}\frac{\left|y_{t}-Y_{t-1}'\theta\right|}{\widehat{g}_{t}}.
\end{flalign*}
To obtain the asymptotic theory of
$\widetilde{\theta}_{an}$, we need two more assumptions.

\begin{ass}
$u_{t}$ is an $\alpha$-mixing process with $\alpha_{u}(k)\leq Ck^{-\delta_{3}}$ for some $C>0$ and $\delta_{3}\geq2(4+\delta_{0})/\delta_{0}$,
where $\alpha_{u}(k)$ is the sequence of strong mixing coefficients of $u_{t}$, and
$\delta_{0}$ is defined as in Assumption \ref{a2.4}(i).
 \label{a6.2}
\end{ass}

\begin{ass}
(i) $K(\cdot)\geq0$ is a bounded continuous function with $\int_{-\infty}^{\infty}K(x)dx=1$;
(ii) $b+1/(nb^{5+\delta_{4}})\to0$ as $n\to\infty$ for some $\delta_{4}>0$.\label{a6.3}
\end{ass}

\noindent Assumption \ref{a6.2} from Shao and Yu (1996) is a technical condition for some moment inequalities
of the partial sums of strong mixing sequences. An exponentially fast decaying
$\alpha_{u}(k)$, which holds for large classes
of processes (see, Carrasco and Chen, 2002), is sufficient for the validity of this assumption.
Assumption \ref{a6.3}(i) holds for commonly used kernels such as uniform, Epanechnikov, biweight, and Gaussian. Assumption \ref{a6.3}(ii) requires that
$b$ converges to zero at a slower rate than $n^{-1/(5+\delta_{4})}$, and this rate can
be improved under some stronger conditions; see Remark \ref{rem_3} below.


The following theorem shows that $\widetilde{\theta}_{an}$ has the same asymptotic variance as
$\widehat{\theta}_{an}$, and hence it is the desired weighted LADE we are looking for.

\begin{thm}\label{thm5}
Suppose Assumptions \ref{a2.1}-\ref{a2.5} and \ref{a6.1}-\ref{a6.3} hold. Then,
\begin{flalign*}
\sqrt{n}(\widetilde{\theta}_{an}-\theta_{0})=\sqrt{n}(\widehat{\theta}_{an}-\theta_{0})+o_{p}(1)\to_{d}N\left(0,
\frac{1}{4}\Sigma_{a2}^{-1}\Sigma_{a1}\Sigma_{a2}^{-1}\right)
\end{flalign*}
as $n\to\infty$, where $\Sigma_{al} (l=1,2)$ is defined in the same way as $\Sigma_{l}$ in (\ref{2.6}) with $w_{t}=g_{t}$.
\end{thm}

\begin{rem}\label{rem_3}
If
$E[|u_{t}||\mathcal{F}_{t-1}]=1$ (a stronger condition than Assumption \ref{a6.1}), we can similarly show that
the result in Theorem \ref{thm5} holds by Propositions \ref{pro_A3}(ii') and \ref{pro_A4}(ii') in Appendix A,
without Assumption \ref{a6.2} and with the replacement of
Assumption \ref{a6.3}(ii) by the weaker condition that $b+1/(nb^{2})\to0$ as $n\to\infty$. In this case,
$b$ can converge to zero at a slower rate than $n^{-1/2}$.
\end{rem}

\begin{rem}
When $\mu_0\equiv0$ and $f_{t}(0)\equiv f(0)$, the asymptotic variance of $\widetilde{\theta}_{an}$ becomes
$\frac{1}{4f^2(0)}\Lambda^{-1}$ by (\ref{simple_version_1}).
\end{rem}

Since the implementation of $\widetilde{\theta}_{an}$ depends on the bandwidth $b$,
we use a similar cross-validation (CV) method as in Xu and Phillips (2008) to select $b$.
That is, $b$ is chosen to be the value of $b^{*}$ which minimizes
\begin{flalign}
\widehat{\mbox{CV}}(b)=\frac{1}{n}\sum_{t=1}^{n}\left(|\widehat{\e}_{t}|-\widehat{g}_{t}\right)^2.\label{cv}
\end{flalign}
In theory, it is unclear whether the bandwidth $b^{*}$ chosen by this CV method
satisfies Assumption \ref{a6.3}(ii). In practice, we can set $b=Cn^{-1/(5+2\delta_{4})}$ for a small
$\delta_4$, and find $C^{*}$ by a grid search such that $b^*=C^*n^{-1/(5+2\delta_{4})}$ minimizes $\widehat{\mbox{CV}}(b)$.
With $b^{*}$ computed by this data-driven method, simulation studies in Section 7 show that
our $\widetilde{\theta}_{an}$ has a good finite sample performance.

Next, we consider the estimation of $g(\cdot)$ as an interest in its own.

\begin{cor}\label{cor6.1}
Suppose (i) Assumptions \ref{a2.1}-\ref{a2.5}, \ref{a6.1}-\ref{a6.2}, and \ref{a6.3}(i) hold; and (ii) $b+1/(nb^2)\to0$ as $n\to\infty$.
Then,
\begin{flalign*}
\plim \limits_{n\to\infty}\widehat{g}_{[n\tau]}=
\left\{
\begin{array}{ll}
g(\tau-)\int_{-\infty}^{0}K(x)dx+g(\tau+)\int_{0}^{\infty}K(x)dx & \mbox{ for }\tau\in(0,1),\\
g(1-)\int_{-\infty}^{0}K(x)dx & \mbox{ for }\tau=1,
\end{array}
\right.
\end{flalign*}
where $g(\tau-)=\lim_{\bar{\tau}\uparrow \tau}g(\bar{\tau})$ and $g(\tau+)=\lim_{\bar{\tau}\downarrow \tau}g(\bar{\tau})$ for $\tau\in(0,1]$.
\end{cor}

Corollary \ref{cor6.1} shows that $\widehat{g}_{[n\tau]}$ converges in probability to $g(\tau)$ for interior points $\tau$ when
$g(\cdot)$ is continuous, but in general to a point between $g(\tau-)$ and $g(\tau+)$ depending on the shape of the kernel function $K(\cdot)$. It is worth noting that the inconsistency of $\widehat{g}_{[n\tau]}$ at points of discontinuities has no impact on the asymptotic theory of
$\widetilde{\theta}_{an}$ as shown in Theorem \ref{thm5}.

Third, since the forms of $g(\cdot)$ and $u_{t}$ are not specified,
we use the RW method as before to
estimate the asymptotic covariance matrix of $\widetilde{\theta}_{an}$. Define
\begin{flalign*}
\widetilde{\theta}_{an}^{*}=\arg\min_{\theta\in\Theta}\sum_{t=1}^{n}w_{t}^{*}\frac{\left|y_{t}-Y_{t-1}'\theta\right|}{\widehat{g}_{t}}.
\end{flalign*}
The following theorem indicates that  the distribution of
$\sqrt{n}(\widetilde{\theta}_{an}-\theta_{0})$ can be approximated
by the resampling distribution of $\sqrt{n}(\widetilde{\theta}_{an}^{*}-\widetilde{\theta}_{an})$.

\begin{thm}\label{thm6}
Suppose Assumption \ref{a3.1} and the conditions in Theorem \ref{thm5} hold. Then, conditional
on $\{y_{t}\}_{t=1}^{n}$,
\begin{flalign*}
\sqrt{n}(\widetilde{\theta}_{an}^{*}-\widetilde{\theta}_{an})\to_{d}N\left(0,
\frac{1}{4}\Sigma_{a2}^{-1}\Sigma_{a1}\Sigma_{a2}^{-1}\right)\,\,\,\,\mbox{ in probability}
\end{flalign*}
as $n\to\infty$, where $\Sigma_{al}$ ($l=1, 2$) is defined as in Theorem \ref{thm5}.
\end{thm}

According to Theorems \ref{thm5} and \ref{thm6}, we can approximate the
asymptotic covariance matrix of $\widetilde{\theta}_{an}$ by $\widetilde{V}_{an}$, where
$\widetilde{V}_{an}$ is calculated in the same way as $\widehat{V}_{wn}$ with
$\widehat{\theta}_{wn}$ and $\widehat{\theta}_{wn}^{*}$
replaced by
$\widetilde{\theta}_{an}$ and $\widetilde{\theta}_{an}^{*}$, respectively.
Then, we can construct another Wald test statistic $W_{an}$ based on $\widetilde{V}_{an}$,  to test the linear
null hypothesis $H_{0}$ in (\ref{3.3}), where
\begin{flalign}\label{6.3}
W_{an}=(\Gamma\widetilde{\theta}_{an}-r)'(\Gamma\widetilde{V}_{an}\Gamma')^{-1}(\Gamma\widetilde{\theta}_{an}-r).
\end{flalign}
If $W_{an}$
is larger than the upper-tailed critical value of $\chi^{2}_{s}$, the null hypothesis $H_{0}$
is rejected. Otherwise, $H_{0}$ is not rejected.

In the end, we consider the sign-based portmanteau test based on $\widetilde{\theta}_{an}$.
Define $\widetilde{r}_{an,k}$ and $\widetilde{r}_{an,k}^{*}$ in the same way as
$\widehat{r}_{wn,k}$ and $\widehat{r}_{wn,k}^{*}$, with $\widehat{\theta}_{wn}$ and $\widehat{\theta}_{wn}^{*}$
replaced by
$\widetilde{\theta}_{an}$ and $\widetilde{\theta}_{an}^{*}$, respectively.
Let $\widetilde{r}_{an}=(\widetilde{r}_{an,1},\widetilde{r}_{an,2},\cdots,\widetilde{r}_{an,M})'$ and $\widetilde{r}_{an}^{*}=(\widetilde{r}_{an,1}^{*},\widetilde{r}_{an,2}^{*},\cdots,\widetilde{r}_{an,M}^{*})'$.
The following theorem indicates that the distribution of
$\sqrt{n}\widetilde{r}_{an}$ can be approximated
by the resampling distribution of $\sqrt{n}(\widetilde{r}_{an}^{*}-\widetilde{r}_{an})$.

\begin{thm}\label{thm7}
Suppose the conditions in Theorem \ref{thm6} hold. Then, if model
(\ref{1.1}) is correctly specified,

(i) $\sqrt{n}\widetilde{r}_{an}\to_{d}N\left(0,
\Sigma_{a4}\Sigma_{a3}\Sigma_{a4}'\right)$ as $n\to\infty$;

(ii) conditional
on $\{y_{t}\}_{t=1}^{n}$, $\sqrt{n}(\widehat{r}_{wn}^{*}-\widehat{r}_{wn})\to_{d}N\left(0,
\Sigma_{a4}\Sigma_{a3}\Sigma_{a4}'\right)$ in probability
as $n\to\infty$, where $\Sigma_{al} (l=3, 4)$ is defined in the same way as $\Sigma_{l}$ in (\ref{4.1}) with $w_{t}=g_{t}$.
\end{thm}

\begin{rem}
When $\mu_0\equiv0$ and $f_{t}(0)\equiv f(0)$, the asymptotic variance of $\widetilde{r}_{an}$ becomes $I_{M}-\ddot{K}_{3}'\Lambda^{-1}\ddot{K}_{3}$ by (\ref{simple_version_2}).
\end{rem}

According to Theorem \ref{thm7}, we can approximate the
asymptotic covariance matrix of $\widetilde{r}_{an}$ by $\widetilde{U}_{an}$, where
$\widetilde{U}_{an}$ is calculated in the same way as $\widehat{U}_{wn}$ with
$\widehat{\theta}_{wn}$ and $\widehat{\theta}_{wn}^{*}$
replaced by
$\widetilde{\theta}_{an}$ and $\widetilde{\theta}_{an}^{*}$, respectively. Then,
we can construct another sign-based portmanteau test statistic $S_{an}(M)$, based on $\widetilde{U}_{an}$,  to detect the adequacy of model (\ref{1.1}), where
\begin{flalign}\label{6.4}
S_{an}(M)=\widetilde{r}_{an}'\widetilde{U}_{an}^{-1}\widetilde{r}_{an}.
\end{flalign}
If $S_{an}(M)$
is larger than the upper-tailed critical value of $\chi^{2}_{M}$, the fitted model (\ref{1.1})
is not adequate. Otherwise, it is adequate.

\section{Simulation study}

In this section, we first assess the performance of the LADE $\widehat{\theta}_{n}$, the feasible ALADE $\widetilde{\theta}_{an}$, and the corresponding RW
approach in finite samples. We generate 1000 replications
of sample size $n=100$ and $200$ from the following model:
\begin{flalign}\label{7.1}
y_{t}=\theta_0 y_{t-1}+\e_{t}\,\,\,\mbox{ and }\,\,\,\e_{t}=g_{t}u_{t},
\end{flalign}
where $\theta_0=0.5$, the error generating process for $\e_{t}$ is designed as follows:
\begin{flalign}\label{7.2}
u_{t}=\eta_{t}\sigma_{t}\,\,\,\mbox{ with }\,\,\,\sigma_{t}^2=0.1+\alpha_{\dag}u_{t-1}^2+\beta_{\dag}\sigma_{t-1}^2;
\end{flalign}
and $g_{t}=g(t/n)$ with $g(\cdot)$ satisfying one of the following  structures:
\begin{flalign}
[\mbox{Abrupt change}]\,\,\,&g(x)=1+(\delta-1)I(x\geq 0.5),\label{7.3}\\
[\mbox{Gradual change}]\,\,\,&g(x)=1+(\delta-1)x^2,\label{7.4}\\
[\mbox{Periodic change}]\,\,\,&g(x)=\sin(\delta x)+2.\label{7.5}
\end{flalign}
Here, we use $(\alpha_{\dag},\beta_{\dag})=(0, 0)$ or $(0.1, 0.8)$, choose $\eta_{t}\sim$ i.i.d. $\mbox{SL}(0, 1)$, $\mbox{N}(0, 1)$ or $\mbox{ST}_{3}$ in model (\ref{7.2}), and
set $\delta\in\{0.2, 5\}$ in models (\ref{7.3})-(\ref{7.4}), $\delta\in\{2\pi, 4\pi\}$ in model (\ref{7.5}). Although $u_{t}$ does not satisfy $E|u_{t}|=1$
in the aforementioned set-up, the estimation property of $\theta_0$ is unaffected by noting that
$\e_{t}=g^{\dag}_{t}u_{t}^{\dag}$ with $g^{\dag}_{t}=g_{t}E|u_{t}|$ and $u_{t}^{\dag}=u_{t}/E|u_{t}|$.
To save space, we only report the results for model (\ref{7.3}) in what follows, and
the results for models (\ref{7.4})-(\ref{7.5}) are similar and can be found in Zhu (2018).

\begin{table}[!h]
\caption{\label{table1}The values of SE and AE of $\widehat{\theta}_{n}$ and $\widetilde{\theta}_{an}$ for model (\ref{7.1})
with $g_{t}\sim$ model (\ref{7.3})}
\scriptsize\addtolength{\tabcolsep}{-2.8pt}
 \renewcommand{\arraystretch}{1.4}
\begin{tabular}{ccc ccccc ccccc ccccc ccc}

\hline

&&&&&&&&&&
\multicolumn{3}{c}{$\eta_{t}\sim \mbox{SL}(0, 1)$}
&&\multicolumn{3}{c}{$\eta_{t}\sim \mbox{ST}_{3}$}
&&\multicolumn{3}{c}{$\eta_{t}\sim \mbox{N}(0, 1)$} \\

\cline{11-13} \cline{15-17} \cline{19-21}

\multicolumn{1}{c}{$\alpha_{\dag}$}&&\multicolumn{1}{c}{$\beta_{\dag}$}&&\multicolumn{1}{c}{$\delta$}&&\multicolumn{1}{c}{$n$}&&
&&\multicolumn{1}{c}{$\widehat{\theta}_{n}$}&&\multicolumn{1}{c}{$\widetilde{\theta}_{an}$}
&&\multicolumn{1}{c}{$\widehat{\theta}_{n}$}&&\multicolumn{1}{c}{$\widetilde{\theta}_{an}$}
&&\multicolumn{1}{c}{$\widehat{\theta}_{n}$}&&\multicolumn{1}{c}{$\widetilde{\theta}_{an}$}\\

\hline

0.0&&0.0&&0.2&&100&&SE &&0.0813&&0.0680&&0.0893&&0.0770&&0.1322&&0.1094\\

&&&&&&&&AE &&0.0898&&0.0758&&0.0924&&0.0800&&0.1288&&0.1118\\

&&&&&&200&& SE&&0.0588& &0.0494&&0.0632&&0.0537&&0.0923&&0.0763\\

&&&&&&&& AE&&0.0616& &0.0520&&0.0643&&0.0551&&0.0913&&0.0783\\

&&&&5&&100&& SE&&0.0897& &0.0759&&0.0942&&0.0833&&0.1352&&0.1159\\

&&&&&& &&AE&&0.0975& &0.0825&&0.1005&&0.0869&&0.1381&&0.1189\\

&&&&&&200&& SE&&0.0564& &0.0473&&0.0615&&0.0540&&0.0930&&0.0768\\

&&&&&& &&AE&&0.0634& &0.0527&&0.0676&&0.0574&&0.0952&&0.0813\\

\cline{1-21}

0.1&&0.8&&0.2&&100&&SE &&0.0962&&0.0786&&0.1020&&0.0867&&0.1304&&0.1109\\

&&&&&&&&AE &&0.1012&&0.0852&&0.1088&&0.0919&&0.1372&&0.1162\\

&&&&&&200&& SE&&0.0631& &0.0539&&0.0767&&0.0623&&0.0951&&0.0822\\

&&&&&&&& AE&&0.0714& &0.0579&&0.0787&&0.0652&&0.0985&&0.0826\\

&&&&5&&100&&  SE&&0.1013& &0.0836&&0.1103&&0.0954&&0.1443&&0.1200\\

&&&&&& &&AE&&0.1103& &0.0911&&0.1174&&0.0984&&0.1489&&0.1256\\

&&&&&&200&& SE&&0.0707& &0.0568&&0.0803&&0.0647&&0.0985&&0.0854\\

&&&&&& &&AE&&0.0731& &0.0599&&0.0806&&0.0667&&0.1056&&0.0882\\

\hline

\end{tabular}
\end{table}

Table \ref{table1} reports the sample root mean squared error (SE) and the average bootstrapped sample root mean squared error (AE) of $\widehat{\theta}_{n}$ and $\widetilde{\theta}_{an}$, based on 1000 replications.
In all calculations, $\widetilde{\theta}_{an}$ is obtained by using Gaussian kernel and the CV method in (\ref{cv}) to choose $b$;
and the bootstrapped sample root mean squared errors for $\widehat{\theta}_{n}$ and $\widetilde{\theta}_{an}$ are computed
by using the RW method with the
bootstrap sample size $J=500$.
From Table \ref{table1},  we can find that
(i) the disparity between SE and AE in each case
is small, indicating that our RW approach is reliable; (ii) the SE of $\widetilde{\theta}_{an}$
is always smaller than the one of $\widehat{\theta}_{n}$ as expected; (iv)
the SE of each estimator becomes small as the sample size $n$ increases;
(v) the SE of each estimator in the case of heavy-tailed $\eta_{t}$ (i.e., $\eta_{t}\sim \mbox{SL}(0, 1)$ or $\mbox{ST}_{3}$) is smaller than
the corresponding one in the case of light-tailed $\eta_{t}$ (i.e., $\eta_{t}\sim \mbox{N}(0, 1)$).
Besides SE, we also compute the sample median and sample median of absolute deviation
of $\widehat{\theta}_{n}$ and $\widetilde{\theta}_{an}$ to see their variability, and the details are relegated to Zhu (2018).

Furthermore, we examine the performance of our CV method by calculating the value of the following ratio:
\begin{flalign*}
R_{1}=\frac{\mbox{SD of } \widetilde{\theta}_{an}}{\mbox{SD of } \widehat{\theta}_{an}},
\end{flalign*}
where SD stands for the sample standard deviation based on 1000 replications.
If our CV method works well, the value of $R_{1}$ should be close to one. Meanwhile, we
compare the efficiency of $\widehat{\theta}_{n}$, $\widetilde{\theta}_{an}$ and the LSE $\widecheck{\theta}_{n}$ with the
infeasible ALSE $\widecheck{\theta}_{an}$ by looking at the values of the following three ratios:
\begin{flalign*}
R_{2}=\frac{\mbox{SD of } \widehat{\theta}_{n}}{\mbox{SD of } \widecheck{\theta}_{an}},\,\,\,
R_{3}=\frac{\mbox{SD of } \widetilde{\theta}_{an}}{\mbox{SD of } \widecheck{\theta}_{an}}
\,\,\,\mbox{ and }\,\,\,
R_{4}=\frac{\mbox{SD of } \widecheck{\theta}_{n}}{\mbox{SD of } \widecheck{\theta}_{an}}.
\end{flalign*}
Clearly, $\widehat{\theta}_{n}$, $\widetilde{\theta}_{an}$ and $\widecheck{\theta}_{n}$ are more efficient than $\widecheck{\theta}_{an}$
when the values of $R_{2}$, $R_{3}$ and $R_{4}$ are smaller than one, respectively.
Also, the most efficient estimator among  $\widehat{\theta}_{n}$, $\widetilde{\theta}_{an}$ and $\widecheck{\theta}_{n}$
is related to the smallest value of $R_{2}$, $R_{3}$ and $R_{4}$.

Table \ref{table2} reports the
values of these four ratios. From this table,
we first find that the value of $R_{1}$ is close to one, and hence it implies that
our CV method works satisfactorily. Second,
we can see that
$\widetilde{\theta}_{an}$ is more efficient
than $\widecheck{\theta}_{an}$ (with values of $R_{3}$ less than one) in general when $\eta_{t}$ is heavy-tailed, and this efficiency advantage in the case of $(\alpha_{\dag},\beta_{\dag})=(0, 0)$
is less substantial than the one in the case of $(\alpha_{\dag},\beta_{\dag})=(0.1, 0.8)$;
on the other hand,  as expected, $\widecheck{\theta}_{an}$ is generally more efficient than
$\widetilde{\theta}_{an}$ when $\eta_{t}$ is light-tailed. For $\widehat{\theta}_{n}$, it can still
be more efficient than $\widecheck{\theta}_{an}$ (with values of $R_{2}$ less than one) in most of
examined cases especially when $\eta_{t}\sim\mbox{SL}(0, 1)$, and this is not the case
when $\eta_{t}\sim\mbox{N}(0, 1)$. Third, we note that $\widecheck{\theta}_{n}$ is always less efficient than
$\widetilde{\theta}_{an}$ and $\widecheck{\theta}_{an}$, and it is more efficient than
$\widehat{\theta}_{n}$ only when $\eta_{t}\sim\mbox{N}(0, 1)$.

\begin{table}[!ht]
\caption{\label{table2}The values of $R_{i}$ ($i=1, 2, 3, 4$) for model (\ref{7.1}) with $g_{t}\sim$ model (\ref{7.3})}
\scriptsize\addtolength{\tabcolsep}{-5.4pt}
 \renewcommand{\arraystretch}{1.5}
\begin{tabular}{ccc ccccc ccccc ccccc ccccc cc cccccc}

\hline

&&&&&&
&&\multicolumn{7}{c}{$\eta_{t}\sim \mbox{SL}(0, 1)$}
&&\multicolumn{7}{c}{$\eta_{t}\sim \mbox{ST}_{3}$}
&&\multicolumn{7}{c}{$\eta_{t}\sim \mbox{N}(0, 1)$} \\

\cline{9-15} \cline{17-23} \cline{25-31}

\multicolumn{1}{c}{$\alpha_{\dag}$}&&\multicolumn{1}{c}{$\beta_{\dag}$}&&\multicolumn{1}{c}{$\delta$}&&\multicolumn{1}{c}{$n$}
&&\multicolumn{1}{c}{$R_{1}$}&&\multicolumn{1}{c}{$R_{2}$}&&\multicolumn{1}{c}{$R_{3}$}&&\multicolumn{1}{c}{$R_{4}$}
&&\multicolumn{1}{c}{$R_{1}$}&&\multicolumn{1}{c}{$R_{2}$}&&\multicolumn{1}{c}{$R_{3}$}&&\multicolumn{1}{c}{$R_{4}$}
&&\multicolumn{1}{c}{$R_{1}$}&&\multicolumn{1}{c}{$R_{2}$}&&\multicolumn{1}{c}{$R_{3}$}&&\multicolumn{1}{c}{$R_{4}$}\\

\hline

0.0&&0.0&&0.2&&100 &&1.0341&&1.0375&&0.8653&&1.4035&&1.0914&&1.1754&&1.0140&&1.4778&&1.0705&&1.6754&&1.3871&&1.5480\\

&&&&&&200&&1.0296& &0.9774&&0.8220&&1.3821&&1.0583&&1.0652&&0.9050&&1.4160&&1.0734&&1.6254&&1.3449&&1.4804\\

&&&&5&&100&&1.0302& &0.9944&&0.8415&&1.2922&&1.0516&&1.1202&&0.9909&&1.3496&&1.0420&&1.5078&&1.2974&&1.3422\\

&&&&&&200&&1.0284& &0.9634&&0.8073&&1.3432&&1.0334&&1.0061&&0.8879&&1.3140&&1.0186&&1.5142&&1.2521&&1.3577\\

\cline{1-31}

0.1&&0.8&&0.2&&100 &&1.0457&&1.0456& &0.8581&&1.3895&&1.0062&&1.0844&&0.9240&&1.3134&&1.0428&&1.5114&&1.2841&&1.4655\\

&&&&&&200&&1.0040& &0.9013&&0.7696&&1.3530&&0.9998&&1.0026&&0.8142&&1.3148&&1.0281&&1.5076&&1.3065&&1.4204\\

&&&&5&&100&&1.0049& &0.9727&&0.8048&&1.2610&&0.9939&&1.0219&&0.8870&&1.2561&&0.9917&&1.4868&&1.2356&&1.3240\\

&&&&&&200&&1.0050& &0.9728&&0.7821&&1.3056&&0.9922&&1.0144&&0.8189&&1.2813&&1.0006&&1.4119&&1.2257&&1.3133\\

\hline

\end{tabular}
\end{table}

In summary, our numerical results in
Tables \ref{table1}-\ref{table2} show that the RW method can provide reliable estimators of the standard deviations for both
$\widehat{\theta}_{n}$ and $\widetilde{\theta}_{an}$, and $\widetilde{\theta}_{an}$ calculated by the CV method shall be recommended
for the heavy-tailed $\eta_{t}$.

Next, we examine the performance of the Wald tests $W_{wn}$ and $W_{an}$, the sign-based portmanteau tests $S_{wn}(M)$ and
$S_{an}(M)$, and the corresponding RW
approach in finite samples. We generate 1000 replications
of sample size $n=100$ and $200$ from the following model:
\begin{flalign}\label{7.6}
y_{t}=\phi_{10}y_{t-1}+\phi_{20} y_{t-2}+\e_{t}\,\,\,\mbox{ and }\,\,\,\e_{t}=g_{t}u_{t},
\end{flalign}
where $(\phi_{10},\phi_{20})=(0.5, \kappa)$ with $\kappa=0, 0.2$ or $0.4$, and
the error generating process for $\e_{t}$ is given by model (\ref{7.1}).
We fit each replication by an AR($2$) model with the LADE (or the feasible ALADE) of
$(\phi_{10},\phi_{20})$, and then apply $W_{wn}$ in (\ref{3.2}) (or $W_{an}$ in (\ref{6.3}))
to test the hypothesis of $\phi_{20}=0$ (i.e., $\Gamma=(0,1)$, $\theta_0=(\phi_{10},\phi_{20})'$ and
 $r=0$ in (\ref{3.3})). Furthermore, we fit each replication by an AR($1$)
model with the LADE (or the feasible ALADE)  of $\phi_{10}$,
and then apply $S_{wn}(M)$ in (\ref{4.3}) (or $S_{an}(M)$ in (\ref{6.4}))
to check whether this fitted AR(1) model is adequate. In all cases, we
set the significance level $\underline{\alpha}=0.05$ and $M=6$.
Based on 1000 replications, the empirical power of all the tests are reported in
Table \ref{table3} for the case that $\eta_{t}\sim\mbox{SL}(0, 1)$, and
their sizes correspond to the results for the case that $\kappa=0$.
Since the
results for other two distributions of $\eta_{t}$ are similar, they are not provided here for saving the space.
From
Table \ref{table3}, it is clear that the sizes of the two Wald tests are close to their nominal ones, and the sizes of
the two portmanteau tests are conservative especially when $n$ is small.
For the power of all
tests, we find that all the power becomes
large as the value of $n$ or $\kappa$
increases; the power of $W_{an}$ and $S_{an}$ based on the feasible ALADE
is larger than that of $W_{wn}$ and $S_{wn}$ based on the LADE, respectively, and this power advantage is
more distinct for the Wald test.
Overall, all tests
based on the
RW approach have a good performance especially
when the sample size is large, and we recommend $W_{an}$ and $S_{an}$ based on the feasible ALADE
in practice.

\begin{table}[!h]
\caption{\label{table3}The power ($\times100$) of all tests for model (\ref{7.6}) with $g_{t}\sim$ model (\ref{7.3}) and $\eta_{t}\sim$ \mbox{SL}(0, 1)}
\scriptsize\addtolength{\tabcolsep}{-4.6pt}
 \renewcommand{\arraystretch}{1.5}
\begin{tabular}{ccc ccccc ccccc ccccc ccccc cccccccc}

\hline

&&&&&&
&&\multicolumn{7}{c}{$\kappa=0$}
&&\multicolumn{7}{c}{$\kappa=0.2$}
&&\multicolumn{7}{c}{$\kappa=0.4$} \\

\cline{9-15} \cline{17-23} \cline{25-31}

\multicolumn{1}{c}{$\alpha_{\dag}$}&&\multicolumn{1}{c}{$\beta_{\dag}$}&&\multicolumn{1}{c}{$\delta$}
&&\multicolumn{1}{c}{$n$}
&&\multicolumn{1}{c}{$W_{wn}$}&&\multicolumn{1}{c}{$W_{an}$}&&\multicolumn{1}{c}{$S_{w}(6)$}&&\multicolumn{1}{c}{$S_{a}(6)$}
&&\multicolumn{1}{c}{$W_{wn}$}&&\multicolumn{1}{c}{$W_{an}$}&&\multicolumn{1}{c}{$S_{w}(6)$}&&\multicolumn{1}{c}{$S_{a}(6)$}
&&\multicolumn{1}{c}{$W_{wn}$}&&\multicolumn{1}{c}{$W_{an}$}&&\multicolumn{1}{c}{$S_{w}(6)$}&&\multicolumn{1}{c}{$S_{a}(6)$}\\

\hline

0.0&&0.0&&0.2&&100 &&4.2&&4.4&&1.8&&2.2&&45.8&&60.6&&8.3&&9.3&&94.5&&97.9&&42.9&&48.8\\

&&&&&&200&&4.6&&4.7&&2.9&&3.1&&76.8&&87.2&&26.5&&27.8&&99.8&&99.9&&87.2&&89.6\\

&&&&5&&100&&3.5&&3.8&&1.3&&2.0&&39.0&&52.3&&7.1&&7.6&&91.2&&96.4&&37.6&&39.3\\

&&&&&&200&&4.7&&4.9&&2.8&&2.8&&71.9&&85.7&&25.2&&25.0&&99.4&&99.9&&85.0&&86.1\\

0.1&&0.8&&0.2&&100 &&3.3&&3.2&&1.7&&1.8&&42.3&&53.5&&6.8&&8.7&&92.7&&98.0&&43.2&&48.6\\

&&&&&&200&&5.2&&4.4&&3.0&&2.9&&66.6&&81.2&&23.3&&24.6&&99.4&&99.9&&84.5&&87.0\\

&&&&5&&100&&4.2&&3.2&&1.6&&1.7&&32.0&&43.9&&6.7&&6.9&&86.2&&94.4&&35.4&&37.8\\

&&&&&&200&&4.7&&4.4&&2.5&&3.2&&59.9&&78.0&&22.0&&22.1&&99.0&&99.8&&83.2&&83.9\\

\cline{1-25}

\hline

\end{tabular}
\end{table}


\section{Concluding remarks}

This paper provides an entire statistical inference procedure for the AR($p$) model
under (conditional) heteroscedasticity of unknown form by
establishing the asymptotic normality of the weighted LADE,
developing the RW method to estimate its asymptotic covariance matrix,
and constructing a portmanteau test for the model diagnostic checking.
This entire procedure can either be based on the usual LADE or the feasible ALADE, and as demonstrated by
the simulation results, the
ALADE is the better choice in terms of the estimation efficiency and testing power. Applications to three U.S. economic data sets
are considered in the supplementary material, and the results
indicate that after the financial crisis in 2007-2008, the monetary policies may become less prudent, and their control on the PPI and CPI
seems to be weaker. The methodology developed in this paper provides a new way to handle the heteroscedastic time
series data and shall have a large applicable scope in practice. Extensions to
other time series models (e.g.,
autoregressive moving average models) and estimation methods (e.g.,
M- and quantile estimations) could be potentially interesting  future works.

\setcounter{equation}{0}
\appendix

\section{Proofs of Theorems}

This appendix gives the proofs of all the theorems, and the proofs of all the corollaries are offered in Zhu (2018).
Define
\begin{flalign}
&z_{n}(\xi)=\frac{1}{\sqrt{n}}\sum_{t=1}^{n}\frac{Y_{t-1}[\mbox{sgn}(\e_{t})]}{\xi_{t}}\label{A1}\\
&\mbox{ and }Z_{n}(\xi)=\sum_{t=1}^{n}\frac{1}{\xi_{t}}\int_{0}^{v'Y_{t-1}/\sqrt{n}}\left\{I(\e_{t}\leq s)-I(\e_{t}\leq 0)\right\}ds,\label{A2}
\end{flalign}
where $v\in\mathcal{R}^{p+1}$. To facilitate the proof of Theorem \ref{thm1}, the following two propositions are needed, and their proofs
are given in Zhu (2018).

\begin{pro}\label{pro_A1}
Suppose Assumptions \ref{a2.1}-\ref{a2.2} and \ref{a2.4}-\ref{a2.5} hold. Then,

(i) $\plim\limits_{n\to\infty}\frac{1}{n}\sum_{t=1}^{n}w_{t}^{-2}Y_{t-1}Y_{t-1}'=\Sigma_{1}$;

(ii) $\plim\limits_{n\to\infty}\frac{1}{n}\sum_{t=1}^{n}(g_{t}w_{t})^{-1}f_{t}(0)Y_{t-1}Y_{t-1}'=\Sigma_{2}$,

\noindent where $\Sigma_{1}$ and $\Sigma_{2}$ are defined as in (\ref{2.6}).
\end{pro}

\begin{pro}\label{pro_A2}
Suppose Assumptions \ref{a2.1}-\ref{a2.4} hold. Then,
$
z_{n}(w)\to_{d} N(0,\Sigma_{1})
$ as $n\to\infty$,
where $z_{n}(w)$ and $\Sigma_{1}$ are defined as in (\ref{A1}) and (\ref{2.6}), respectively.
\end{pro}

\textsc{Proof of Theorem \ref{thm1}.}
Denote
$H_{n}(v)=\sum_{t=1}^{n}\frac{1}{w_{t}}\left(\left|\e_{t}-\frac{v'}{\sqrt{n}}Y_{t-1}\right|-|\e_{t}|\right)$,
where $v\in\mathcal{R}^{p+1}$. Then, $\widehat{v}_{wn}:=\sqrt{n}(\widehat{\theta}_{wn}-\theta_{0})$ is the
the minimizer of $H_{n}(v)$ over $\mathcal{R}^{p+1}$.

Using the identity
\begin{flalign}
|x-y|-|x|=-y\left\{\mbox{sgn}(x)\right\}+2\int_{0}^{y}\left\{I(x\leq s)-I(x\leq 0)\right\}ds,\label{A11}
\end{flalign}
which holds when $x\not=0$ (see Knight (1998)), it follows that
\begin{flalign}
H_{n}(v)&=-v'z_{n}(w)+2Z_{n}(w),
\label{A13}
\end{flalign}
where $z_{n}(w)$ and $Z_{n}(w)$ are defined as in (\ref{A1}) and (\ref{A2}), respectively.
Write
$Z_{n}(w):=\sum_{t=1}^{n}K_{t}=\sum_{t=1}^{n}E(K_{t}|\mathcal{F}_{t-1})+\sum_{t=1}^{n}\left[K_{t}-E(K_{t}|\mathcal{F}_{t-1})\right].
$
Subsequently, we can show that for each $v\in\mathcal{R}^{p+1}$,
\begin{flalign}
&\sum_{t=1}^{n}E(K_{t}|\mathcal{F}_{t-1})
=v'\frac{\Sigma_{2}}{2}v+o_{p}(1),\label{A19}\\
&\sum_{t=1}^{n}\left[K_{t}-E(K_{t}|\mathcal{F}_{t-1})\right]=o_{p}(1);\label{A18}
\end{flalign}
see Zhu (2018) for details. By (\ref{A13})-(\ref{A18}), $H_{n}(v)=-v'z_{n}(w)+v'\Sigma_{2}v+o_{p}(1)$  for each $v\in\mathcal{R}^{p+1}$.
As $H_{n}(v)$ is convex for each $v$,
Theorem 2 in Kato (2009) implies that
\begin{flalign}
\widehat{v}_{wn}=(2\Sigma_{2})^{-1}z_{n}(w)+o_{p}(1).\label{A22}
\end{flalign}
Now, the conclusion follows from (\ref{A22}) and Proposition \ref{pro_A2}.
\hfill$\square$

\vspace{3mm}

\textsc{Proof of Theorem \ref{thm2}.}
Denote $H_{n}^{*}(v)=\sum_{t=1}^{n}\frac{w_{t}^{*}}{w_{t}}\left(\left|\e_{t}-\frac{v'}{\sqrt{n}}Y_{t-1}\right|-|\e_{t}|\right)$,
where $v\in\mathcal{R}^{p+1}$. Then, $\widehat{v}_{wn}^{*}=\sqrt{n}(\widehat{\theta}_{wn}^{*}-\theta_{0})$ is the minimizer of $H_{n}^{*}(v)$ over $\mathcal{R}^{p+1}$.

As for (\ref{A13}), we can obtain that
$H_{n}^{*}(v)=-v'z_{n}^{*}(w)+2Z_{n}^{*}(w)$,
where
\begin{flalign*}
&z_{n}^{*}(w)=\frac{1}{\sqrt{n}}\sum_{t=1}^{n}\frac{w_{t}^{*}Y_{t-1}[\mbox{sgn}(\e_{t})]}{w_{t}}\\
&\mbox{ and }Z_{n}^{*}(w)=\sum_{t=1}^{n}\frac{w_{t}^{*}}{w_{t}}\int_{0}^{v'Y_{t-1}/\sqrt{n}}\left\{I(\e_{t}\leq s)-I(\e_{t}\leq 0)\right\}ds.
\end{flalign*}
Since $\{w_{t}^{*}\}$ is independent of $\{y_{t}\}$ with $E(w_{t}^{*})=1$ by Assumption \ref{a3.1},
a similar argument as for (\ref{A19})-(\ref{A18}) entails that
$Z_{n}^{*}(w)=v'\frac{\Sigma_{2}}{2}v+o_{p}(1)$. Then, for each $v\in\mathcal{R}^{p+1}$,
$
H_{n}^{*}(v)=-v'z_{n}^{*}(w)+v'\Sigma_{2}v+o_{p}(1)
$,
which implies that
\begin{flalign}\label{A25}
\widehat{v}_{wn}^{*}=(2\Sigma_{2})^{-1}z_{n}^{*}(w)+o_{p}(1)
\end{flalign}
by a similar argument as for (\ref{A22}). Hence, by (\ref{A22}) and (\ref{A25}), we have
\begin{flalign}\label{A26}
\sqrt{n}(\widehat{\theta}_{wn}^{*}-\widehat{\theta}_{wn})&=
\frac{[2\Sigma_{2}]^{-1}}{\sqrt{n}}\sum_{t=1}^{n}\frac{(w_{t}^{*}-1)Y_{t-1}\mbox{sgn}(\e_{t})}{w_{t}}+o_{p}(1)\nonumber\\
&=:[2\Sigma_{2}]^{-1}\sum_{t=1}^{n}J_{tn}+o_{p}(1).
\end{flalign}

Let $E^{*}$ be the conditional
expectation on $\{y_{t}\}_{t=1}^{n}$ and $\lambda\in\mathcal{R}^{p+1}$ be a non-zero constant vector.
We now study the conditional distribution of $\sum_{t=1}^{n}\lambda'J_{tn}$.
First, since $\{w_{t}^{*}\}$ is independent of $\{y_{t}\}$ with $Ew_{t}^{*}=1$, we have
\begin{flalign}\label{A27}
E^{*}[\lambda'J_{tn}]=0.
\end{flalign}
Next, since $\mbox{var}(w_{t}^{*})=1$, by the independence of $\{w_{t}^{*}\}$ and $\{y_{t}\}$, we have
\begin{flalign}\label{A28}
\sum_{t=1}^{n}E^{*}\left[\lambda'J_{tn}J_{tn}'\lambda\right]&=
\lambda'\left\{\frac{1}{n}\sum_{t=1}^{n}\frac{Y_{t-1}Y_{t-1}'[\mbox{sgn}(\e_{t})]^{2}}{w_{t}^2}\right\}\lambda\nonumber\\
&=\lambda'\Sigma_{1}\lambda+o_{p}(1)
\end{flalign}
by Proposition \ref{pro_A1}(i). Finally, we claim that the
Lindeberg condition holds by showing that
\begin{flalign}\label{A29}
&\sum_{t=1}^{n}E^{*}\left[\lambda'J_{tn}J_{tn}'\lambda I(|\lambda'J_{tn}|>\eta)\right]=o_{p}(1);
\end{flalign}
see Zhu (2018) for details. By (\ref{A27})-(\ref{A29}), the Cram\'{e}r-Wold device and the central limit theorem in
Pollard (1984, Theorem VIII.1) yield that conditional on $\{y_{t}\}_{t=1}^{n}$, $\sum_{t=1}^{n}J_{tn}\to_{d} N(0,\Sigma_{1})\mbox{ in probability as } n\to\infty$.
Now, the conclusion follows directly from (\ref{A26}).
\hfill$\square$

\vspace{3mm}


\textsc{Proof of Theorem \ref{thm3}.}
By Lemma B.4(i) with $s_{n}=\widehat{v}_{wn}$, we have
\begin{flalign}\label{A30}
\sqrt{n}\widehat{r}_{wn,k}=\frac{1}{\sqrt{n}}\sum_{t=k+1}^{n}\mbox{sgn}(\e_{t}(\widehat{\theta}_{wn}))
\mbox{sgn}(\e_{t-k}(\widehat{\theta}_{wn}))+o_{p}(1).
\end{flalign}
Rewrite $\mbox{sgn}(\e_{t}(\widehat{\theta}_{wn}))
\mbox{sgn}(\e_{t-k}(\widehat{\theta}_{wn}))-\mbox{sgn}(\e_{t})
\mbox{sgn}(\e_{t-k})=P_{nt}(\widehat{v}_{wn})+Q_{nt}(\widehat{v}_{wn})$,
where $P_{nt}(v)$ and $Q_{nt}(v)$ are defined as in Lemma B.4.
By (\ref{A30}) and Lemma B.4(ii)-(iv), it follows that
\begin{flalign}\label{A31}
\sqrt{n}\widehat{r}_{wn,k}=\left[\frac{1}{\sqrt{n}}\sum_{t=k+1}^{n}\mbox{sgn}(\e_{t})
\mbox{sgn}(\e_{t-k})\right]-2\Omega_{3k}'\widehat{v}_{wn}+o_{p}(1).
\end{flalign}
Hence, by (\ref{A22}) and (\ref{A31}), we have
\begin{flalign}\label{A32}
\sqrt{n}\widehat{r}_{wn}=\Sigma_{4}S_{n}+o_{p}(1),
\end{flalign}
where
$S_{n}=\frac{1}{\sqrt{n}}\big(\sum_{t=2}^{n}\mbox{sgn}(\e_{t})
\mbox{sgn}(\e_{t-1}), \cdots, \sum_{t=M+1}^{n}\mbox{sgn}(\e_{t})
\mbox{sgn}(\e_{t-M}),\sum_{t=1}^{n}$ $\frac{Y_{t-1}'\mbox{sgn}(\e_{t})}{w_{t}}\big)'$.
Finally, the conclusion holds by the central limit theorem for m.d.s. and a similar argument as
for Proposition \ref{pro_A1}. \hfill$\square$

\vspace{3mm}

\textsc{Proof of Theorem \ref{thm4}.}
By Assumption \ref{a3.1} and Lemma B.4(i) with $s_{n}=\widehat{v}_{wn}^{*}$, we have
\begin{flalign*}
\sqrt{n}\widehat{r}_{wn,k}^{*}=\frac{1}{\sqrt{n}}\sum_{t=k+1}^{n}w_{t}^{*}\mbox{sgn}(\e_{t}(\widehat{\theta}_{wn}^{*}))
\mbox{sgn}(\e_{t-k}(\widehat{\theta}_{wn}^{*}))+o_{p}(1).
\end{flalign*}
By Assumption \ref{a3.1} and a similar argument as for (\ref{A31}), we can obtain that
\begin{flalign}\label{A33}
\sqrt{n}\widehat{r}_{wn,k}^{*}=\left[\frac{1}{\sqrt{n}}\sum_{t=k+1}^{n}w_{t}^{*}\mbox{sgn}(\e_{t})
\mbox{sgn}(\e_{t-k})\right]-2\Omega_{3k}'\widehat{v}_{wn}^{*}+o_{p}(1).
\end{flalign}
Hence, by (\ref{A25}) and (\ref{A33}), we have
\begin{flalign}\label{A34}
\sqrt{n}\widehat{r}_{wn}^{*}=\Sigma_{4}S_{n}^{*}+o_{p}(1),
\end{flalign}
where $S_{n}^{*}=\frac{1}{\sqrt{n}}\big(\sum_{t=2}^{n}w_{t}^{*}\mbox{sgn}(\e_{t})
\mbox{sgn}(\e_{t-1}), \cdots, \sum_{t=M+1}^{n}w_{t}^{*}\mbox{sgn}(\e_{t})
\mbox{sgn}(\e_{t-M}),$ $\sum_{t=1}^{n}\frac{w_{t}^{*}Y_{t-1}'\mbox{sgn}(\e_{t})}{w_{t}}\big)'$.
Finally, the conclusion holds by (\ref{A32}), (\ref{A34}) and a similar argument as
for Theorem \ref{thm2}. \hfill$\square$

\vspace{3mm}

Following a similar terminology as in  Robinson (1987), we let
\begin{flalign*}
\widetilde{g}_{t}=\sum_{i=1}^{n}k_{ti}|\e_{i}|\mbox{ and }
\overline{g}_{t}=\sum_{i=1}^{n}k_{ti}g_{i}.
\end{flalign*}
To prove Theorems \ref{thm5}-\ref{thm7}, two crucial propositions are needed, and their
proofs are given in Zhu (2018).

\begin{pro}\label{pro_A3}
Suppose Assumptions \ref{a2.1}-\ref{a2.5}, \ref{a6.1}-\ref{a6.2} and \ref{a6.3}(i) hold. Then,

(i) $z_{n}(\widehat{g})-z_{n}(\widetilde{g})=O_{p}\left(\frac{1}{\sqrt{n}b}\right)$;

(ii) $z_{n}(\widetilde{g})-z_{n}(\overline{g})
=O_{p}\left(\frac{1}{\sqrt{n}b^{\frac{5+\delta_{4}}{2}}}+\frac{1}{nb^{3+\delta_{4}}}+b^{\delta_{4}}\right)\mbox{ for some }\delta_{4}>0;
$

(ii') $z_{n}(\widetilde{g})-z_{n}(\overline{g})=O_{p}\left(\frac{1}{\sqrt{n}b}\right)$ if $E[|u_{t}||\mathcal{F}_{t-1}]=1$;

(iii) $z_{n}(\overline{g})-z_{n}(g)=o_{p}(1)$,

\noindent where $z_{n}(\cdot)$ is defined as in (\ref{A1}).
\end{pro}

\begin{pro}\label{pro_A4}
Suppose the conditions in Proposition \ref{pro_A3} hold. Then, for each $v\in\mathcal{R}^{p+1}$,

(i) $Z_{n}(\widehat{g})-Z_{n}(\widetilde{g})=O_{p}\left(\frac{1}{n^{3/4}b}\right)$;

(ii) $Z_{n}(\widetilde{g})-Z_{n}(\overline{g})=O_{p}\left(\frac{1}{n^{1/4}b}\right)$;

(ii') $Z_{n}(\widetilde{g})-Z_{n}(\overline{g})=O_{p}\left(\frac{1}{n^{1/4}b^{1/2}}\right)$ if $E[|u_{t}||\mathcal{F}_{t-1}]=1$;

(iii) $Z_{n}(\overline{g})-Z_{n}(g)=o_{p}(1)$,

\noindent where $Z_{n}(\cdot)$ is defined as in (\ref{A2}).
\end{pro}

\textsc{Proof of Theorem \ref{thm5}.}  Denote
$\widehat{G}_{n}(v)=\sum_{t=1}^{n}\frac{1}{\widehat{g}_{t}}\left(\left|\e_{t}-\frac{v'}{\sqrt{n}}Y_{t-1}\right|-|\e_{t}|\right)$,
where $v\in\mathcal{R}^{p+1}$. Then, $\widetilde{v}_{an}:=\sqrt{n}(\widetilde{\theta}_{an}-\theta_{0})$ is
the minimizer of $\widehat{G}_{n}(v)$ over $\mathcal{R}^{p+1}$.
As for (\ref{A13}),
$
\widehat{G}_{n}(v)=-v'z_{n}(\widehat{g})+2Z_{n}(\widehat{g}).
$
By Assumption \ref{a6.3}(ii) and Propositions \ref{pro_A3}-\ref{pro_A4}, it follows that for each $v\in\mathcal{R}^{p+1}$,
$
\widehat{G}_{n}(v)=-v'z_{n}(g)+2Z_{n}(g)+o_{p}(1).
$
Hence, as for (\ref{A22}), we can deduce that
\begin{flalign}\label{A38}
\widetilde{v}_{an}=(2\Sigma_{a2})^{-1}z_{n}(g)+o_{p}(1),
\end{flalign}
and the conclusion follows by Proposition \ref{pro_A2}. \hfill$\square$

\vspace{3mm}

\textsc{Proof of Theorem \ref{thm6}.}  Denote $
\widehat{G}_{n}^{*}(v)=\sum_{t=1}^{n}\frac{w_{t}}{\widehat{g}_{t}}\left(\left|\e_{t}-\frac{v'}{\sqrt{n}}Y_{t-1}\right|-|\e_{t}|\right)$,
where $v\in\mathcal{R}^{p+1}$. Then,
$\widetilde{v}_{an}^{*}:=\sqrt{n}(\widetilde{\theta}_{an}^{*}-\theta_{0})$ is
the minimizer of $\widehat{G}_{n}^{*}(v)$ over $\mathcal{R}^{p+1}$.
As for (\ref{A13}), $\widehat{G}_{n}^{*}(v)=-v'z_{n}^{*}(\widehat{g})+2Z_{n}^{*}(\widehat{g})$.
By Assumptions \ref{a3.1} and \ref{a6.3}(ii) and similar arguments as for Propositions \ref{pro_A3}-\ref{pro_A4}, it follows that for each $v\in\mathcal{R}^{p+1}$,
$
\widehat{G}_{n}^{*}(v)=-v'z_{n}^{*}(g)+2Z_{n}^{*}(g)+o_{p}(1).
$ Hence, as for (\ref{A25}), we can deduce that
\begin{flalign}\label{A39}
\widetilde{v}_{an}^{*}=(2\Sigma_{a2})^{-1}z_{n}^{*}(g)+o_{p}(1).
\end{flalign}
Now, the conclusion holds by (\ref{A38})-(\ref{A39}) and the similar arguments as for Theorem \ref{thm2}. \hfill$\square$

\vspace{3mm}

\textsc{Proof of Theorem \ref{thm7}.} By noting that $
\widetilde{v}_{an}=O_{p}(1)$ and $\widetilde{v}_{an}^{*}=O_{p}(1)$, the proofs of
(i) and (ii) follow the same arguments as for Theorems \ref{thm3} and \ref{thm4}, respectively.
\hfill$\square$

\section{Some technical lemmas}


This appendix gives Lemmas B.1-B.6, which can be found in Zhu (2018).

\section*{Acknowledgements}
The author greatly appreciates the very helpful comments and suggestions
of anonymous referees, the Associate Editor and the Co-Editor.


\end{document}